\documentclass[12pt]{article}
\pdfoutput=1

\usepackage{amsmath}
\usepackage{amssymb}
\usepackage{epsfig}
\usepackage{epstopdf}
\usepackage{latexsym}
\usepackage{color}
\numberwithin{equation}{section}
\usepackage[
      colorlinks=false,
      linkcolor=darkblue,  
      urlcolor=blue,    
      filecolor=blue,     
      citecolor=red,
linktocpage=true,
      pdfstartview=FitV,
      bookmarksopen=true    
      ]{hyperref}

\begin{document}

\begin{titlepage}

\centerline{\Huge \rm } 
\bigskip
\centerline{\Huge \rm Supersymmetric $AdS_6$ black holes}
\bigskip
\centerline{\Huge \rm from matter coupled $F(4)$ gauged supergravity}
\bigskip
\bigskip
\bigskip
\bigskip
\bigskip
\centerline{\rm Minwoo Suh}
\bigskip
\centerline{\it Department of Physics, Kyungpook National University, Daegu 41566, Korea}
\bigskip
\centerline{\tt minwoosuh1@gmail.com} 
\bigskip
\bigskip
\bigskip
\bigskip
\bigskip
\bigskip
\bigskip
\bigskip
\bigskip
\bigskip
\bigskip
\bigskip

\begin{abstract}
\noindent In matter coupled $F(4)$ gauged supergravity in six dimensions, we study supersymmetric $AdS_6$ black holes with various horizon geometries. We find new $AdS_2\,\times\,\Sigma_{\mathfrak{g}_1}\times\Sigma_{\mathfrak{g}_2}$ horizons with $\mathfrak{g}_1>1$ and $\mathfrak{g}_2>1$, and present the black hole solution numerically. The full black hole is an interpolating geometry between the asymptotically $AdS_6$ boundary and the $AdS_2\,\times\,\Sigma_{\mathfrak{g}_1}\times\Sigma_{\mathfrak{g}_2}$ horizon. We also find black holes with horizons of K\"ahler four-cycles in Calabi-Yau fourfolds and Cayley four-cycles in $Spin(7)$ manifolds.
\end{abstract}

\vskip 7cm

\flushleft {October, 2018}

\end{titlepage}

\tableofcontents

\section{Introduction and conclusions}

In $F(4)$ gauged supergravity in six dimensions \cite{Romans:1985tw}, there is a unique supersymmetric fixed point which is dual to 5d superconformal $USp(2N)$ gauge theory \cite{Seiberg:1996bd, Intriligator:1997pq}. As it was shown in \cite{Cvetic:1999un}, $F(4)$ gauged supergravity is a consistent truncation of massive type IIA supergravity \cite{Romans:1985tz}. The fixed point uplifts to $AdS_6\times_wS^4$ near-horizon geometry of the D4-D8 brane system \cite{Ferrara:1998gv, Brandhuber:1999np}. In the spirit of \cite{Maldacena:2000mw}, supergravity solutions of wrapped D4-branes on various supersymmetric cycles were studied in $F(4)$ gauged supergravity. D4-branes wrapped on two- and three-cycles were studied in \cite{Nunez:2001pt}. They found $AdS_4$ and $AdS_3$ fixed point solutions. See \cite{Dibitetto:2018iar} for more recent results.

Recently, by considering D4-branes wrapped on supersymmetric four-cycles, we found supersymmetric $AdS_6$ black holes of $F(4)$ gauged supergravity in \cite{Suh:2018tul}.{\footnote {Previously, D4-branes wrapped on supersymmetric four-cycles were studied in \cite{Naka:2002jz}, but the equations and solutions were incorrect.}} To be specific, we found the full black hole solutions which is an interpolating geometry between the asymptotically $AdS_6$ boundary and the $AdS_2\,\times\,H^2\,\times\,H^2$ horizon. Via the AdS/CFT correspondence, \cite{Maldacena:1997re}, analogous to the $AdS_4$ black hole cases in \cite{Cacciatori:2009iz, Benini:2015noa, Benini:2015eyy}, the Bekenstein-Hawking entropy of the black holes nicely matched with the topologically twisted index of 5d $USp(2N)$ gauge theory on $\Sigma_{\mathfrak{g}_1}\times\Sigma_{\mathfrak{g}_2}\times{S}^1$ in the large $N$ limit \cite{Hosseini:2018uzp, Crichigno:2018adf}. We also considered black hole horizons of K\"ahler four-cycles in Calabi-Yau fourfolds and Cayley four-cycles in $Spin(7)$ manifolds.

Pure $F(4)$ gauged supergravity is a consistent truncation of massive type IIA supergravity \cite{Cvetic:1999un} and type IIB supergravity \cite{Jeong:2013jfc, Hong:2018amk, Malek:2018zcz} on a four-hemisphere. Although it is not known whether it is also a consistent truncation of ten-dimensional supergravity, one can couple vector multiplets to pure $F(4)$ gauged supergravity \cite{Andrianopoli:2001rs}. In this theory, new fixed points and holographic RG flows were studied in \cite{Karndumri:2012vh, Karndumri:2014lba, Karndumri:2015eta}. See \cite{Karndumri:2016ruc, Gutperle:2017nwo, Gutperle:2018axv} also for other studies in this theory.

In this paper, in matter coupled $F(4)$ gauged supergravity, we continue our study on supersymmetric $AdS_6$ black holes. We consider $F(4)$ gauged supergravity coupled to three vector multiplets, and its $U(1)\,\times\,U(1)$-invariant truncation first considered in \cite{Karndumri:2015eta}. We consider black hole solutions with a horizon which is a product of two Riemann surfaces, $AdS_2\,\times\,\Sigma_{\mathfrak{g}_1}\times\Sigma_{\mathfrak{g}_2}$. We derive supersymmetry equations and obtain $AdS_2$ solutions which was first found in \cite{Hosseini:2018usu}. The $AdS_2$ horizon exists only for the $H^2\,\times\,H^2$ background, and not for the $H^2\,\times\,S^2$ or $S^2\,\times\,S^2$ backgrounds. We present the full black hole solutions numerically.

We also consider black holes with horizons of K\"ahler four-cycles in Calabi-Yau fourfolds and Cayley four-cycles in $Spin(7)$ manifolds. For Cayley four-cycles in $Spin(7)$ manifolds, we consider the $SU(2)_{diag}$-invariant truncation of $F(4)$ gauged supergravity coupled to three vector multiplets. We find new $AdS_2$ horizons. It will be interesting to have a field theory interpretation of this $AdS_2$ solution.

In section 2, we review matter coupled $F(4)$ gauged supergravity in six dimensions. In section 3, we consider $F(4)$ gauged supergravity coupled to three vector multiplets, and its $U(1)\,\times\,U(1)$-invariant truncation. We consider supersymmetric black hole solutions with a horizon which is a product of two Riemann surfaces. In section 4, we consider supersymmetric black hole solutions with horizons of K\"ahler four-cycles in Calabi-Yau fourfolds and Cayley four-cycles in $Spin(7)$ manifolds. In appendix A, we present the equations of motion for the $U(1)\,\times\,U(1)$-invariant truncation.

\bigskip
\bigskip

\noindent {\bf Note added:} In the final stage of this work, we became aware of \cite{Hosseini:2018usu} which has some overlap with the results presented here in section 3 and section 4.1.

\section{Matter coupled $F(4)$ gauged supergravity}

We review matter coupled $F(4)$ gauged supergravity in six dimensions \cite{Andrianopoli:2001rs}. The gravity multiplet consists of 
\begin{equation}
\left(e^a_\mu,\,\psi^A_\mu,\,A^\alpha_\mu,\,B_{\mu\nu},\,\chi^A,\,\sigma\right)\,,
\end{equation}
where they denote the graviton, gravitini, four vector fields, a two-form gauge potential, dilatini, and a real scalar field, respectively. The vector fields, $A^\alpha_\mu$, $\alpha\,=\,0,\,1,\,2,\,3$, can be used to gauge the $SU(2)_R\,\times\,U(1)$ gauge symmetry. The vector multiplet consists of
\begin{equation}
\left(A_\mu,\,\lambda_A,\,\varphi^\alpha\right)^I\,,
\end{equation}
where they denote a vector field, gaugini, and four real scalar fields, respectively, and $I\,=\,1,\ldots,\,n$ labels the vector multiplets. The fermionic fields are eight-dimensional pseudo-Majorana spinors and transform in the fundamental representation of the $SU(2)_R\,\sim\,USp(2)_R$ R-symmetry denoted by indices, $A,B=1,2$. We denote the coupling constants of gauge fields from the gravity and vector multiplets by $g_1$ and $g_2$, respectively, and the mass parameter of the two-form gauge potential by $m$. 

When there is no vector multiplet, the theory reduces to pure $F(4)$ gauged supergravity \cite{Romans:1985tw}. In pure $F(4)$ gauged supergravity, there are five inequivalent theories : $\mathcal{N}\,=\,4^+$ ($g_1>0$, $m>0$), $\mathcal{N}\,=\,4^-$ ($g_1<0$, $m>0$), $\mathcal{N}\,=\,4^g$ ($g_1>0$, $m=0$), $\mathcal{N}\,=\,4^m$ ($g_1=0$, $m>0$), $\mathcal{N}\,=\,4^0$ ($g_1=0$, $m=0$). The $\mathcal{N}\,=\,4^+$ theory admits a supersymmetric $AdS_6$ fixed point when $g_1\,=\,3m$.  At the supersymmetric $AdS_6$ fixed point, all the fields are vanishing except the $AdS_6$ metric.

The scalar fields from the gravity and vector multiplets parametrize each factor of the coset manifold, 
\begin{equation}
\mathcal{M}\,=\,SO(1,1)\times\frac{SO(4,n)}{SO(4){\times}SO(n)}\,,
\end{equation}
respectively. The coset representative of the second factor is given by
\begin{equation}
P^I\,_\alpha\,=\,(P^I\,_0,P^I\,_r)\,=\,(\Omega^I\,_0,\Omega^I\,_r)\,,
\end{equation}
where we define 
\begin{equation} \label{structure}
\Omega^\Lambda\,_\Sigma\,=\,(L^{-1})^\Lambda\,_\Pi\nabla{L}^\Pi\,_\Sigma\,, \qquad \nabla{L}^\Lambda\,_\Sigma\,=\,dL^\Lambda\,_\Sigma-f_\Gamma\,^\Lambda\,_\Pi{A}^\Gamma{L}^\Pi\,_\Sigma\,,
\end{equation}
and $\Lambda=(\alpha,I)$ or $\Lambda=(0,r,I)$ with $r\,=\,1,\,2,\,3$. The indices, $\Lambda$, $\Sigma$, \ldots, are raised and lowered by
\begin{equation}
\eta_{\Lambda\Sigma}\,=\,\text{diag}(1,1,1,1,-1,\ldots,-1)\,.
\end{equation}

The bosonic Lagrangian is given by
\begin{align}
e^{-1}\mathcal{L}\,=\,&-\frac{1}{4}R+\partial_\mu\sigma\partial^\mu\sigma-\frac{1}{4}P_{I\alpha\mu}P^{I\alpha\mu}-V-\frac{1}{8}e^{-2\sigma}\mathcal{N}_{\Lambda\Sigma}F^\Lambda_{\mu\nu}F^{\Sigma\mu\nu}+\frac{3}{64}e^{4\sigma}H_{\mu\nu\lambda}H^{\mu\nu\lambda} \notag \\
&-\frac{1}{64}\epsilon^{\mu\nu\rho\sigma\tau\kappa}B_{\mu\nu}\left(\eta_{\Lambda\Sigma}F^\Lambda_{\rho\sigma}F^\Sigma_{\tau\kappa}+mB_{\rho\sigma}F^0_{\tau\kappa}+\frac{1}{3}m^2B_{\rho\sigma}B_{\tau\kappa}\right)\,,
\end{align}
and
\begin{equation}
V\,=\,-e^{2\sigma}\left(\frac{1}{36}A^2+\frac{1}{4}B^tB_t+\frac{1}{4}C^I\,_tC_{It}+D^I\,_tD_{It}\right)+m^2e^{-6\sigma}\mathcal{N}_{00}-me^{-2\sigma}\left(\frac{2}{3}AL_{00}-2B^tL_{0t}\right)\,,
\end{equation}
where we define
\begin{equation}
\mathcal{N}_{\Lambda\Sigma}\,=\,L^0\,_\Lambda(L^{-1})_{0\Sigma}+L_\Lambda\,^r(L^{-1})_{r\Sigma}-L_\Lambda\,^I(L^{-1})_{I\Sigma}\,,
\end{equation}
and
\begin{equation}
A\,=\,\epsilon^{rst}K_{rst}\,, \qquad B^t\,=\,\epsilon^{trs}K_{rs0}\,, \qquad C_I\,^t\,=\,\epsilon^{trs}K_{rIs}\,, \qquad D_{It}\,=\,K_{0It}\,.
\end{equation}
The field strengths are collectively defined by{\footnote {The numerical factors are due to the unusual convention of form fields, $e.g.$, $\omega\,=\,\omega_{\mu\nu}dx^{\mu}dx^\nu$, in \cite{Andrianopoli:2001rs}.}
\begin{equation}
F^\Lambda_{\mu\nu}\,=\,\partial_{[\mu}A^\Lambda_{\nu]}-m\delta_0^{\Lambda}B_{\mu\nu}\,, \qquad H_{\mu\nu\lambda}\,=\,\partial_{[\mu}{B}_{\nu\lambda]}\,.
\end{equation}

The supersymmetry variations of the fermionic fields are given by{\footnote {There is a missing $\gamma_7$ in (4.32) and also in the $m$ dependent term in (4.35) of \cite{Andrianopoli:2001rs}.}}
\begin{align}
\delta\psi_{\mu{A}}\,=&\,D_\mu\epsilon_A+\frac{i}{24}\left[Ae^\sigma+6me^{-3\sigma}(L^{-1})_{00}\right]\gamma_\mu\epsilon_A-\frac{i}{8}\left[B_te^\sigma-2me^{-3\sigma}(L^{-1})_{t0}\right]\gamma^7\gamma_\mu\sigma^t_{AB}\epsilon^B \notag \\
+&\frac{1}{16}e^{-\sigma}\left[(L^{-1})_{0\Lambda}\left(F^\Lambda_{\nu\lambda}-mB_{\nu\lambda}\delta_0^\Lambda\right)\gamma_7\epsilon_{AB}-(L^{-1})_{r\Lambda}F^\Lambda_{\nu\lambda}\sigma^r_{AB}\right]\left(\gamma_\mu\,^{\nu\lambda}-6\delta_\mu^\nu\gamma^\lambda\right)\epsilon^B \notag \\
+&\frac{i}{32}e^{2\sigma}H_{\nu\lambda\rho}\gamma_7\gamma^{\nu\lambda\rho}\gamma_\mu\epsilon_A\,, \\
\delta\chi_A\,=&\,\frac{i}{2}\gamma^\mu\partial_\mu\sigma\epsilon_A+\frac{1}{24}\left[Ae^\sigma-18me^{-3\sigma}(L^{-1})_{00}\right]\epsilon_A+\frac{1}{8}\left[B_te^\sigma+6me^{-3\sigma}(L^{-1})_{t0}\right]\gamma^7\sigma^t_{AB}\epsilon^B \notag \\
+&\frac{i}{16}e^{-\sigma}\left[(L^{-1})_{0\Lambda}\left(F^\Lambda_{\mu\nu}-mB_{\mu\nu}\delta_0^\Lambda\right)\gamma_7\epsilon_{AB}+(L^{-1})_{r\Lambda}F^\Lambda_{\mu\nu}\sigma^r_{AB}\right]\gamma^{\mu\nu}\epsilon^B \notag \\
+&\frac{1}{32}e^{2\sigma}H_{\mu\nu\lambda}\gamma_7\gamma^{\mu\nu\lambda}\epsilon_A\,, \\
\delta\lambda^I_A\,=&\,iP^I_{\mu{r}}\gamma^\mu\sigma^r_{AB}\epsilon^B-iP^I_{\mu{0}}\gamma^7\gamma^\mu\epsilon_A-e^\sigma\left[C^I\,_t-2i\gamma^7D^I\,_t\right]\sigma^t_{AB}\epsilon^B-2me^{-3\sigma}(L^{-1})^I\,_0\gamma^7\epsilon_A \notag \\
+&\frac{i}{2}e^{-\sigma}(L^{-1})^I\,_\Lambda{F}^\Lambda_{\mu\nu}\gamma^{\mu\nu}\epsilon_A\,,
\end{align}
where we define
\begin{align}
K_{rst}\,&=\,g_1\epsilon_{lmn}L^l\,_r(L^{-1})_s\,^mL^n\,_t+g_2C_{IJK}L^I\,_r(L^{-1})_s\,^JL^K\,_t\,, \notag \\
K_{rs0}\,&=\,g_1\epsilon_{lmn}L^l\,_r(L^{-1})_s\,^mL^n\,_0+g_2C_{IJK}L^I\,_r(L^{-1})_s\,^JL^K\,_0\,, \notag \\
K_{rIt}\,&=\,g_1\epsilon_{lmn}L^l\,_r(L^{-1})_I\,^mL^n\,_t+g_2C_{LJK}L^L\,_r(L^{-1})_I\,^JL^K\,_t\,, \notag \\
K_{0It}\,&=\,g_1\epsilon_{lmn}L^l\,_0(L^{-1})_I\,^mL^n\,_t+g_2C_{LJK}L^L\,_0(L^{-1})_I\,^JL^K\,_t\,,
\end{align}
and
\begin{equation}
D_\mu\,=\,\partial_\mu\epsilon_A+\frac{1}{4}\omega_\mu\,^{ab}\gamma_{ab}\epsilon_A+\frac{i}{2}\left(\frac{1}{2}g_1\epsilon^{rst}\Omega_{\mu{st}}+i\gamma_7\Omega_{\mu{r0}}\right)\sigma_{rAB}\epsilon^B\,.
\end{equation}

The Pauli matrices, $\sigma^{tA}\,_B$, satisfy the relations,
\begin{equation}
\sigma^t_{AB}\,=\,\sigma^{tC}\,_B\epsilon_{CA}\,,
\end{equation}
and $\sigma^t_{AB}\,=\,\sigma^t_{(AB)}$. We also define the chirality matrix by
\begin{equation}
\gamma_7\,=\,i\gamma_0\gamma_1\gamma_2\gamma_3\gamma_4\gamma_5\,,
\end{equation}
with $\gamma_7^2\,=\,-1$ and $\gamma_7^T\,=\,-\gamma_7$. We employ the mostly minus signature, $(+-----)$.

\subsection{$F(4)$ gauged supergravity coupled to three vector multiplets}

In $F(4)$ gauged supergravity coupled to three vector multiplets, we have a scalar field from the gravity multiplet and four scalar fields from each vector multiplet: total thirteen scalar fields. The scalar fields from the gravity multiplet and the vector multiplets parametrize each factor of the coset manifold, respectively,
\begin{equation}
\mathcal{M}\,=\,SO(1,1)\times\frac{SO(4,3)}{SO(4){\times}SO(3)}\,.
\end{equation}
There are three vector fields from the gravity multiplet and three vector fields from the three vector multiplets. Two sets of three vector fields can be used to gauge $SU(2)_R\,\times\,SU(2)$ gauge group. We denote the coupling constants of two $SU(2)$ factors by $g_1$ and $g_2$. The structure constant in \eqref{structure} splits into
\begin{equation}
f_{rst}\,=\,g_1\epsilon_{rst}\,, \qquad f_{IJK}\,=\,g_2C_{IJK}\,=\,g_2\epsilon_{IJK}\,.
\end{equation}
The generators of $SO(4)$, $SU(2)_R$, $SU(2)$, and the non-compact $SO(4,3)$ could be represented by, respectively \cite{Karndumri:2015eta},
\begin{align} \label{generators}
J^{\alpha\beta}\,=\,e^{\beta,\alpha}-e^{\alpha,\beta}, \,\,\,\,\,\,\,\,  J_1^{rs}\,=\,e^{s,r}-e^{r,s}\,, \,\,\,\,\,\,\,\, J_2^{IJ}\,=\,e^{J+3,I+3}-e^{I+3,J+3}\,, \,\,\,\,\,\,\,\, Y_{\alpha{I}}\,=\,e^{\alpha,I+3}+e^{I+3,\alpha}\,,
\end{align}
where we define
\begin{equation}
\left(e^{\Lambda\Sigma}\right)_{\Gamma\Pi}\,=\,\delta_{\Lambda\Gamma}\delta_{\Sigma\Pi}\,.
\end{equation}

\section{Black holes with $AdS_2\,\times\,\Sigma_{\mathfrak{g}_1}\times\Sigma_{\mathfrak{g}_2}$ horizon}

\subsection{The $U(1){\times}U(1)$-invariant truncation}

We truncate the theory to the $U(1){\times}U(1)$-invariant sector, which was first considered in section 3.1 of \cite{Karndumri:2015eta}. The $U(1){\times}U(1)$ are generated by $J^{12}_1$ and $J^{12}_2$. We find two non-compact $SO(4,3)$ generators, $Y_{03}$ and $Y_{33}$, which are invariant under the action of $J^{12}_1$ and $J^{12}_2$. We exponentiate the non-compact generators and obtain the coset representative,
\begin{equation}
L\,=\,e^{\varphi_1Y_{03}}e^{\varphi_2Y_{33}}\,.
\end{equation}
We also have two $U(1)$ gauge fields, $A^3$ and $A^6$, and a two-form gauge potential, $B_{\mu\nu}$, in the $U(1){\times}U(1)$-invariant truncation. The Lagrangian of the truncation is given by
\begin{align} \label{lagrangian}
e^{-1}\mathcal{L}\,=\,&-\frac{1}{4}R+\partial_\mu\sigma\partial^\mu\sigma+\frac{1}{4}\cosh^2\varphi_2\partial_\mu\varphi_1\partial^\mu\varphi_1+\frac{1}{4}\partial_\mu\varphi_2\partial^\mu\varphi_2-V \notag \\ &-\frac{1}{8}e^{-2\sigma}\cosh(2\varphi_2)F^3_{\mu\nu}F^{3\mu\nu}-\frac{1}{8}e^{-2\sigma}\left(\cosh^2\varphi_1\cosh(2\varphi_2)+\sinh^2\varphi_1\right)F^6_{\mu\nu}F^{6\mu\nu} \notag \\ 
&+\frac{1}{4}e^{-2\sigma}\cosh\varphi_1\sinh(2\varphi_2)F^3_{\mu\nu}F^{6\mu\nu}-\frac{1}{8}m^2e^{-2\sigma}B_{\mu\nu}B^{\mu\nu}+\frac{3}{64}e^{4\sigma}H_{\mu\nu\rho}H^{\mu\nu\rho} \notag \\
&-\frac{1}{64}\epsilon^{\mu\nu\rho\sigma\tau\kappa}B_{\mu\nu}\left(F^3_{\rho\sigma}F^3_{\tau\kappa}-F^6_{\rho\sigma}F^6_{\tau\kappa}+\frac{1}{3}m^2B_{\rho\sigma}B_{\tau\kappa}\right)\,,
\end{align}
where the scalar potential is
\begin{equation}
V\,=\,-g_1^2e^{2\sigma}-4g_1me^{-2\sigma}\cosh\varphi_1\cosh\varphi_2+m^2e^{-6\sigma}\left(\cosh^2\varphi_1+\sinh^2\varphi_1\cosh(2\varphi_2)\right)\,.
\end{equation}

\subsection{The supersymmetry equations}

In this section, we obtain supersymmetric $AdS_6$ black holes with a horizon which is a product of two Riemann surfaces. We consider the metric,
\begin{equation}
ds^2\,=\,e^{2F(r)}\left(dt^2-dr^2\right)-e^{2G_1(r)}\left(d\theta_1^2+\sin^2\theta_1d\phi_1^2\right)-e^{2G_2(r)}\left(d\theta_2^2+\sin^2\theta_1d\phi_2^2\right)\,,
\end{equation}
for the $S^2\,\times\,S^2$ background, and
\begin{equation}
ds^2\,=\,e^{2F(r)}\left(dt^2-dr^2\right)-e^{2G_1(r)}\left(d\theta_1^2+\sinh^2\theta_1d\phi_1^2\right)-e^{2G_2(r)}\left(d\theta_2^2+\sinh^2\theta_1d\phi_2^2\right)\,,
\end{equation}
for the $H^2\,\times\,H^2$ background. The only non-vanishing components of the non-Abelian $SU(2)$ gauge field, $A^\Lambda_\mu$, $\Lambda\,=\,0,\,1,\,\ldots,\,6$, are given by
\begin{equation}
A^3\,=\,-2a_1\cos\theta_1{d}\phi_1-2a_2\cos\theta_2{d}\phi_2\,, \qquad A^6\,=\,-2b_1\cos\theta_1{d}\phi_1-2b_2\cos\theta_2{d}\phi_2\,,
\end{equation}
for the $S^2\,\times\,S^2$ background and
\begin{equation}
A^3\,=\,2a_1\cosh\theta_1{d}\phi_1+2a_2\cosh\theta_2{d}\phi_2\,, \qquad A^6\,=\,2b_1\cosh\theta_1{d}\phi_1+2b_2\cosh\theta_2{d}\phi_2\,,
\end{equation}
for the $H^2\,\times\,H^2$ background, where the magnetic charges, $a_1$, $a_2$, $b_1$, $b_2$, are constant. In order to have equal signs for the field strengths, we set opposite signs of the gauge fields for the $S^2\,\times\,S^2$ and $H^2\,\times\,H^2$ backgrounds. We also have a non-trivial two-form gauge potential, $B_{\mu\nu}$, determined by solving the equations of motion,
\begin{equation}
B_{tr}\,=\,-\frac{1}{2m^2}\left(a_1a_2-b_1b_2\right)e^{2\sigma+2F-2G_1-2G_2}\,.
\end{equation}
The three-form field strength of the two-form gauge potential, $H_{\mu\nu\lambda}$, vanishes identically.

The supersymmetry equations are obtained by setting the supersymmetry variations of the fermionic fields to zero. From the supersymmetry variations, we obtain
\begin{align} \label{oneone}
F'e^{-F}\gamma^{\hat{r}}\epsilon_A&+\frac{i}{2}\left(g_1e^\sigma\cosh\varphi_2+me^{-3\sigma}\cosh\varphi_1\right)\epsilon_A-\frac{i}{2}me^{-3\sigma}\sinh\varphi_1\sinh\varphi_2\gamma_7\sigma^3_{AB}\epsilon^B \notag \\
&+\frac{1}{4}e^{-\sigma}\sinh\varphi_1\left(b_1e^{-2G_1}\gamma^{\hat{\theta}_1\hat{\phi}_1}+b_2e^{-2G_2}\gamma^{\hat{\theta}_2\hat{\phi}_2}\right)\gamma_7\epsilon_A \notag \\ 
&-\frac{1}{4}e^{-\sigma}\cosh\varphi_2\left(a_1e^{-2G_1}\gamma^{\hat{\theta}_1\hat{\phi}_1}+a_2e^{-2G_2}\gamma^{\hat{\theta}_2\hat{\phi}_2}\right)\sigma^3_{AB}\epsilon^B \notag \\ 
&+\frac{1}{4}e^{-\sigma}\cosh\varphi_1\sinh\varphi_2\left(b_1e^{-2G_1}\gamma^{\hat{\theta}_1\hat{\phi}_1}+b_2e^{-2G_2}\gamma^{\hat{\theta}_2\hat{\phi}_2}\right)\sigma^3_{AB}\epsilon^B \notag \\ 
&+\frac{3}{8m}\left(a_1a_2-b_1b_2\right)e^{\sigma-2G_1-2G_2}\gamma^{\hat{t}\hat{r}}\gamma_7\epsilon_A\,=\,0\,,
\end{align}
\begin{align} \label{twotwo}
G_1'e^{-F}\gamma^{\hat{r}}\epsilon_A&+\frac{i}{2}\left(g_1e^\sigma\cosh\varphi_2+me^{-3\sigma}\cosh\varphi_1\right)\epsilon_A-\frac{i}{2}me^{-3\sigma}\sinh\varphi_1\sinh\varphi_2\gamma_7\sigma^3_{AB}\epsilon^B \notag \\
&-\frac{1}{4}e^{-\sigma}\sinh\varphi_1\left(3b_1e^{-2G_1}\gamma^{\hat{\theta}_1\hat{\phi}_1}-b_2e^{-2G_2}\gamma^{\hat{\theta}_2\hat{\phi}_2}\right)\gamma_7\epsilon_A \notag \\ 
&+\frac{1}{4}e^{-\sigma}\cosh\varphi_2\left(3a_1e^{-2G_1}\gamma^{\hat{\theta}_1\hat{\phi}_1}-a_2e^{-2G_2}\gamma^{\hat{\theta}_2\hat{\phi}_2}\right)\sigma^3_{AB}\epsilon^B \notag \\ 
&-\frac{1}{4}e^{-\sigma}\cosh\varphi_1\sinh\varphi_2\left(3b_1e^{-2G_1}\gamma^{\hat{\theta}_1\hat{\phi}_1}-b_2e^{-2G_2}\gamma^{\hat{\theta}_2\hat{\phi}_2}\right)\sigma^3_{AB}\epsilon^B \notag \\ 
&-\frac{1}{8m}\left(a_1a_2-b_1b_2\right)e^{\sigma-2G_1-2G_2}\gamma^{\hat{t}\hat{r}}\gamma_7\epsilon_A\,=\,0\,,
\end{align}
\begin{align} \label{threethree}
G_2'e^{-F}\gamma^{\hat{r}}\epsilon_A&+\frac{i}{2}\left(g_1e^\sigma\cosh\varphi_2+me^{-3\sigma}\cosh\varphi_1\right)\epsilon_A-\frac{i}{2}me^{-3\sigma}\sinh\varphi_1\sinh\varphi_2\gamma_7\sigma^3_{AB}\epsilon^B \notag \\
&-\frac{1}{4}e^{-\sigma}\sinh\varphi_1\left(3b_2e^{-2G_2}\gamma^{\hat{\theta}_1\hat{\phi}_1}-b_1e^{-2G_1}\gamma^{\hat{\theta}_2\hat{\phi}_2}\right)\gamma_7\epsilon_A \notag \\ 
&+\frac{1}{4}e^{-\sigma}\cosh\varphi_2\left(3a_2e^{-2G_2}\gamma^{\hat{\theta}_1\hat{\phi}_1}-a_1e^{-2G_1}\gamma^{\hat{\theta}_2\hat{\phi}_2}\right)\sigma^3_{AB}\epsilon^B \notag \\ 
&-\frac{1}{4}e^{-\sigma}\cosh\varphi_1\sinh\varphi_2\left(3b_2e^{-2G_2}\gamma^{\hat{\theta}_1\hat{\phi}_1}-b_1e^{-2G_1}\gamma^{\hat{\theta}_2\hat{\phi}_2}\right)\sigma^3_{AB}\epsilon^B \notag \\ 
&-\frac{1}{8m}\left(a_1a_2-b_1b_2\right)e^{\sigma-2G_1-2G_2}\gamma^{\hat{t}\hat{r}}\gamma_7\epsilon_A\,=\,0\,,
\end{align}
\begin{align} \label{fourfour}
\sigma'e^{-F}\gamma^{\hat{r}}\epsilon_A&-\frac{i}{2}\left(g_1e^\sigma\cosh\varphi_2-3me^{-3\sigma}\cosh\varphi_1\right)\epsilon_A-\frac{3i}{2}me^{-3\sigma}\sinh\varphi_1\sinh\varphi_2\gamma_7\sigma^3_{AB}\epsilon^B \notag \\
&-\frac{1}{4}e^{-\sigma}\sinh\varphi_1\left(b_1e^{-2G_1}\gamma^{\hat{\theta}_1\hat{\phi}_1}+b_2e^{-2G_2}\gamma^{\hat{\theta}_2\hat{\phi}_2}\right)\gamma_7\epsilon_A \notag \\ 
&+\frac{1}{4}e^{-\sigma}\cosh\varphi_2\left(a_1e^{-2G_1}\gamma^{\hat{\theta}_1\hat{\phi}_1}+a_2e^{-2G_2}\gamma^{\hat{\theta}_2\hat{\phi}_2}\right)\sigma^3_{AB}\epsilon^B \notag \\ 
&-\frac{1}{4}e^{-\sigma}\cosh\varphi_1\sinh\varphi_2\left(b_1e^{-2G_1}\gamma^{\hat{\theta}_1\hat{\phi}_1}+b_2e^{-2G_2}\gamma^{\hat{\theta}_2\hat{\phi}_2}\right)\sigma^3_{AB}\epsilon^B \notag \\ 
&+\frac{1}{8m}\left(a_1a_2-b_1b_2\right)e^{\sigma-2G_1-2G_2}\gamma^{\hat{t}\hat{r}}\gamma_7\epsilon_A\,=\,0\,,
\end{align}
\begin{align} \label{fivefive}
\varphi_2'e^{-F}\gamma^{\hat{r}}\epsilon_A&-\varphi_1'e^{-F}\cosh\varphi_2\gamma_7\gamma^{\hat{r}}\sigma^3_{AB}\epsilon^B-2ig_1e^\sigma\sinh\varphi_2\epsilon_A-2ime^{-3\sigma}\sinh\varphi_1\cosh\varphi_2\gamma_7\sigma^3_{AB}\epsilon^B \notag \\
&-e^{-\sigma}\sinh\varphi_2\left(a_1e^{-2G_1}\gamma^{\hat{\theta}_1\hat{\phi}_1}+a_2e^{-2G_2}\gamma^{\hat{\theta}_2\hat{\phi}_2}\right)\sigma^3_{AB}\epsilon^B \notag \\ 
&+e^{-\sigma}\cosh\varphi_1\cosh\varphi_2\left(b_1e^{-2G_1}\gamma^{\hat{\theta}_1\hat{\phi}_1}+b_2e^{-2G_2}\gamma^{\hat{\theta}_2\hat{\phi}_2}\right)\sigma^3_{AB}\epsilon^B\,=\,0\,,
\end{align}
where the hatted indices are the flat indices. The $t$-, $\theta_1$-, and $\theta_2$-components of the gravitino variations give \eqref{oneone}, \eqref{twotwo}, \eqref{threethree}, the dilatino variation gives \eqref{fourfour}, and the gaugino variation gives \eqref{fivefive}. The $\phi_1$-, $\phi_2$-components of the gravitino variations are identical to the $\theta_1$-, and $\theta_2$-components beside few more terms,
\begin{equation} \label{pretwist}
\epsilon_A\,=\,-2ig_1a_1\gamma^{\hat{\theta}_1\hat{\phi}_1}\sigma^3_{AB}\epsilon^B\,, \qquad \epsilon_A\,=\,-2ig_1a_2\gamma^{\hat{\theta}_2\hat{\phi}_2}\sigma^3_{AB}\epsilon^B\,.
\end{equation}

We employ the projection conditions,
\begin{equation} \label{proj}
\gamma^{\hat{r}}\epsilon_A\,=\,i\epsilon_A\,, \qquad \gamma^{\hat{\theta}_1\hat{\phi}_1}\sigma^3_{AB}\epsilon^B\,=\,-i\lambda\epsilon_A\,, \qquad \gamma^{\hat{\theta}_2\hat{\phi}_2}\sigma^3_{AB}\epsilon^B\,=\,-i\lambda\epsilon_A\,, 
\end{equation}
where $\lambda\,=\,\pm1$. Solutions with the projection conditions preserve $1/8$ of the supersymmetries. As we check later, in order to have consistent supersymmetry equations with the equations of motion, it is required to have
\begin{equation} \label{phi1zero}
\varphi_1\,=\,0\,.
\end{equation}
Therefore, from now on, we set $\varphi_1$ to vanish. 

We present the complete supersymmetry equations,
\begin{align} \label{susy11}
F'e^{-F}\,=\,&-\frac{1}{2}\left(g_1e^\sigma\cosh\varphi_2+me^{-3\sigma}\right)-\frac{3}{8m}\left(a_1a_2-b_1b_2\right)e^{\sigma-2G_1-2G_2} \notag \\ &-\frac{\lambda}{4}e^{-\sigma}\cosh\varphi_2\left(a_1e^{-2G_1}+a_2e^{-2G_2}\right)+\frac{\lambda}{4}e^{-\sigma}\sinh\varphi_2\left(b_1e^{-2G_1}+b_2e^{-2G_2}\right)\,, \notag \\
G_1'e^{-F}\,=\,&-\frac{1}{2}\left(g_1e^\sigma\cosh\varphi_2+me^{-3\sigma}\right)+\frac{1}{8m}\left(a_1a_2-b_1b_2\right)e^{\sigma-2G_1-2G_2} \notag \\ &+\frac{\lambda}{4}e^{-\sigma}\cosh\varphi_2\left(3a_1e^{-2G_1}-a_2e^{-2G_2}\right)-\frac{\lambda}{4}e^{-\sigma}\sinh\varphi_2\left(3b_1e^{-2G_1}-b_2e^{-2G_2}\right)\,, \notag \\
G_2'e^{-F}\,=\,&-\frac{1}{2}\left(g_1e^\sigma\cosh\varphi_2+me^{-3\sigma}\right)+\frac{1}{8m}\left(a_1a_2-b_1b_2\right)e^{\sigma-2G_1-2G_2} \notag \\ &+\frac{\lambda}{4}e^{-\sigma}\cosh\varphi_2\left(3a_2e^{-2G_2}-a_1e^{-2G_1}\right)-\frac{\lambda}{4}e^{-\sigma}\sinh\varphi_2\left(3b_2e^{-2G_2}-b_1e^{-2G_1}\right)\,, \notag \\
\sigma'e^{-F}\,=\,&+\frac{1}{2}\left(g_1e^\sigma\cosh\varphi_2-3me^{-3\sigma}\right)-\frac{1}{8m}\left(a_1a_2-b_1b_2\right)e^{\sigma-2G_1-2G_2} \notag \\ &+\frac{\lambda}{4}e^{-\sigma}\cosh\varphi_2\left(a_1e^{-2G_1}+a_2e^{-2G_2}\right)-\frac{\lambda}{4}e^{-\sigma}\sinh\varphi_2\left(b_1e^{-2G_1}+b_2e^{-2G_2}\right)\,, \notag \\
\varphi_2'e^{-F}\,=\,&+2g_1e^\sigma\sinh\varphi_2-{\lambda}e^{-\sigma}\sinh\varphi_2\left(a_1e^{-2G_1}+a_2e^{-2G_2}\right)+{\lambda}e^{-\sigma}\cosh\varphi_2\left(b_1e^{-2G_1}+b_2e^{-2G_2}\right)\,.
\end{align}
We also obtain twist conditions on the magnetic charges from \eqref{pretwist},
\begin{equation} \label{twist}
a_1\,=\,-\frac{k}{2{\lambda}g_1}\,, \qquad a_2\,=\,-\frac{k}{2{\lambda}g_1}\,.
\end{equation}
where $k\,=\,+1$ for the $S^2\,\times\,S^2$ background and $k\,=\,-1$ for the $H^2\,\times\,H^2$ background.{\footnote {It is possible to have geometries like $S^2\times{H}^2$ for $k_1\,=\,+1$ and $k_2\,=\,-1$, or vice versa. One can easily generalize our supersymmetry equations and the twist conditions to that case.}} There is no condition on $b_1$ and $b_2$. The supersymmetry equations are consistent with the equations of motion. We present the equations of motion in appendix A.

\subsection{The $AdS_2$ solutions}

In this section, we find $AdS_2$ solutions of the supersymmetry equations. The solutions describe the $AdS_2\,\times\,\Sigma_{\mathfrak{g}_1}\times\Sigma_{\mathfrak{g}_2}$ horizon of six-dimensional black holes. 

Now  we will consider the $\mathcal{N}\,=\,4^+$ theory, $g_1>0$, $m>0$. When $b_1\,=\,b_2\,=\,0$, we find an $AdS_2$ fixed point solution for the $H^2\,\times\,H^2$ background with $k\,=\,-1$,
\begin{equation} \label{ads2ads2}
e^{F}\,=\,\frac{1}{2^{5/4}g_1^{3/4}m^{1/4}}\frac{1}{r}\,, \qquad e^{G_1}\,=\,e^{G_2}\,=\,\frac{1}{2^{3/4}g_1^{3/4}m^{1/4}}\,, \qquad e^{\sigma}\,=\,\frac{2^{1/4}m^{1/4}}{g_1^{1/4}}\,, \qquad e^{\varphi_2}\,=\,1\,,
\end{equation}
which is the $AdS_2$ solution first found in \cite{Suh:2018tul}.{\footnote{In order to compare with \cite{Suh:2018tul}, we have to reparametrize our parameters by
\begin{equation} \label{reparam}
\sigma\,\rightarrow\,\frac{1}{\sqrt{2}}\phi\,, \qquad g_1,\,m \,\rightarrow\,\frac{1}{2\sqrt{2}}g,\,\frac{1}{2\sqrt{2}}m\,, \qquad a_1,\,a_2\,\rightarrow\,\sqrt{2}a_1,\,\sqrt{2}a_2\,.
\end{equation}}}
When we consider the $S^2\,\times\,S^2$ background with $k\,=\,+1$, $AdS_2$ fixed point does not exist.
When we consider for non-zero $b_1$ and $b_2$, we obtain the $AdS_2$ solutions, first found in \cite{Hosseini:2018usu},
\begin{align} \label{ads2sol}
e^F\,=&\,\frac{e^{-\sigma}}{2g_1\cosh\varphi_2}\frac{1}{r}\,, \notag \\
e^{2G_1}\,=&\,\frac{\lambda}{2m}e^{2\sigma}(a_1\cosh\varphi_2-b_1\sinh\varphi_2)\,, \notag \\
e^{2G_2}\,=&\,\frac{\lambda}{2m}e^{2\sigma}(a_2\cosh\varphi_2-b_2\sinh\varphi_2)\,, \notag \\
e^{4\sigma}\,=&\,\frac{m}{2}\frac{2(a_1a_2-b_1b_2)+(a_1a_2+b_1b_2)\cosh(2\varphi_2)-(a_1b_2+a_2b_1)\sinh(2\varphi_2)}{g_1\cosh\varphi_2(a_1\cosh\varphi_2-b_1\sinh\varphi_2)(a_2\cosh\varphi_2-b_2\sinh\varphi_2)}\,, \notag \\
e^{2\varphi_2}\,=&\,\frac{1}{(a_1-b_1)(a_2-b_2)}\Big[\left(a_1a_2+a_1b_2+a_2b_1-3b_1b_2\right)-2(\Phi/2)^{1/3} \notag \\
&-\left.\frac{2}{(\Phi/2)^{1/3}}\left(a_1a_2^2b_1+a_1^2a_2b_2-a_1a_2b_1b_2-2a_2b_1^2b_2-2a_1b_1b_2^2+3b_1^2b_2^2\right)\right]\,,
\end{align}
where we define
\begin{align}
\Phi\,=\,&-\left(a_1a_2-2b_1b_2\right)\left(a_1^2b_2^2+a_2^2b_1^2\right)-\left(a_1b_2+a_2b_1\right)\left(a_1^2a_2^2+10b_1^2b_2^2-7a_1a_2b_1b_2\right) \notag \\
&-2b_1b_2\left(a_1^2a_2^2-5b_1^2b_2^2+a_1a_2b_1b_2\right) \notag \\
&+\sqrt{\left(a_1-b_1\right)^2\left(a_2-b_2\right)^2\left(a_1a_2-2b_1b_2\right)\left(a_1^2b_2+a_1a_2b_1-2b_1^2b_2\right)\left(a_2^2b_1+a_1a_2b_2-2b_1b_2^2\right)}\,.
\end{align}
All the fields are parametrized by the magnetic charges, $(a_1,a_2,b_1,b_2)$. As $(a_1,a_2)$ are fixed by the twist condition in \eqref{twist}, there are two free parameters left, $(b_1,b_2)$.

In order to have $AdS_2$ solutions, we should choose $(b_1,b_2)$ which makes
\begin{equation}
e^F>0\,, \qquad e^{2G_1}>0,\, \qquad e^{2G_2}>0,\, \qquad e^{4\sigma}>0,\, \qquad e^{2\varphi_2}>0\,.
\end{equation}
We plot the range of $(b_1,b_2)$ which satisfies the positivity conditions for $H^2\,\times\,H^2$, $H^2\,\times\,S^2$, $S^2\,\times\,H^2$ and $S^2\,\times\,S^2$, respectively. The positivity ranges are depicted in figure 1. We set $m\,=\,1/2$ and $g_1\,=\,3m$ to have a unit radius for the $AdS_6$ boundary. {\it From the plots, we conjecture that only the $H^2\,\times\,H^2$ background gives the $AdS_2$ solutions.}{\footnote {We note that, when $G_1\,\leftrightarrow\,G_2$, the solutions are invariant under $a_1\,\leftrightarrow\,a_2$ and $b_1\,\leftrightarrow\,b_2$. However, as $a_1\,=\,a_2$ from \eqref{twist}, they are invariant under $b_1\,\leftrightarrow\,b_2$. When we plot the positivity range for $H^2\,\times\,S^2$ and $S^2\,\times\,H^2$, there are small and irregular distributions of points. Most of the points do not respect the invariance under $b_1\,\leftrightarrow\,b_2$. Even for the points invariant under $b_1\,\leftrightarrow\,b_2$ seem not to give $AdS_2$ solutions. We presume that the appearance of these irregular distribution in the positivity range would be due to the complexity of the conditions.}} Even in the large region in the graph for the $H^2\,\times\,H^2$ background, only a small part near origin yields $AdS_2$ solutions.
\begin{figure}[h!]
\begin{center}
\includegraphics[width=2in]{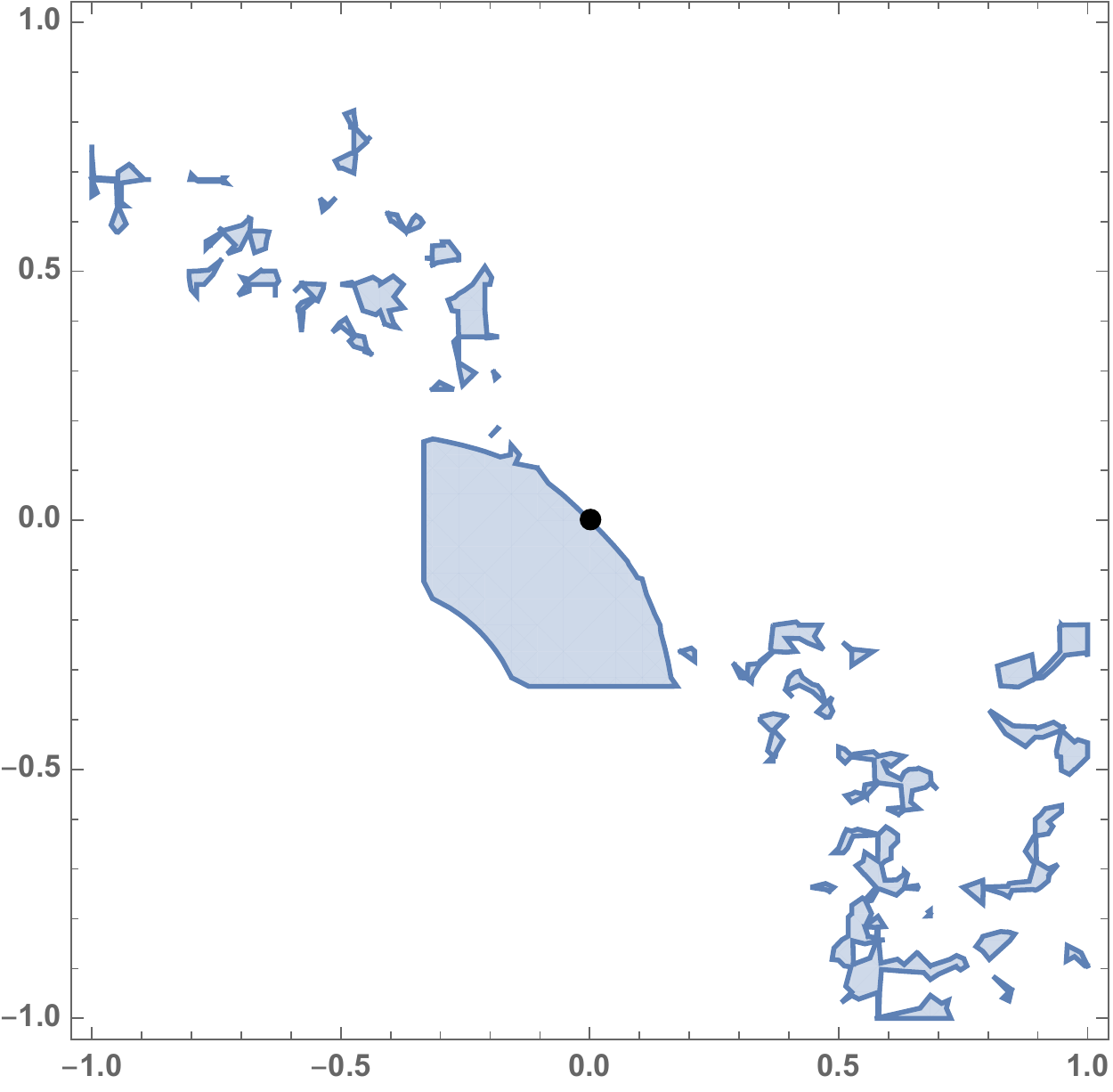} \qquad \includegraphics[width=2in]{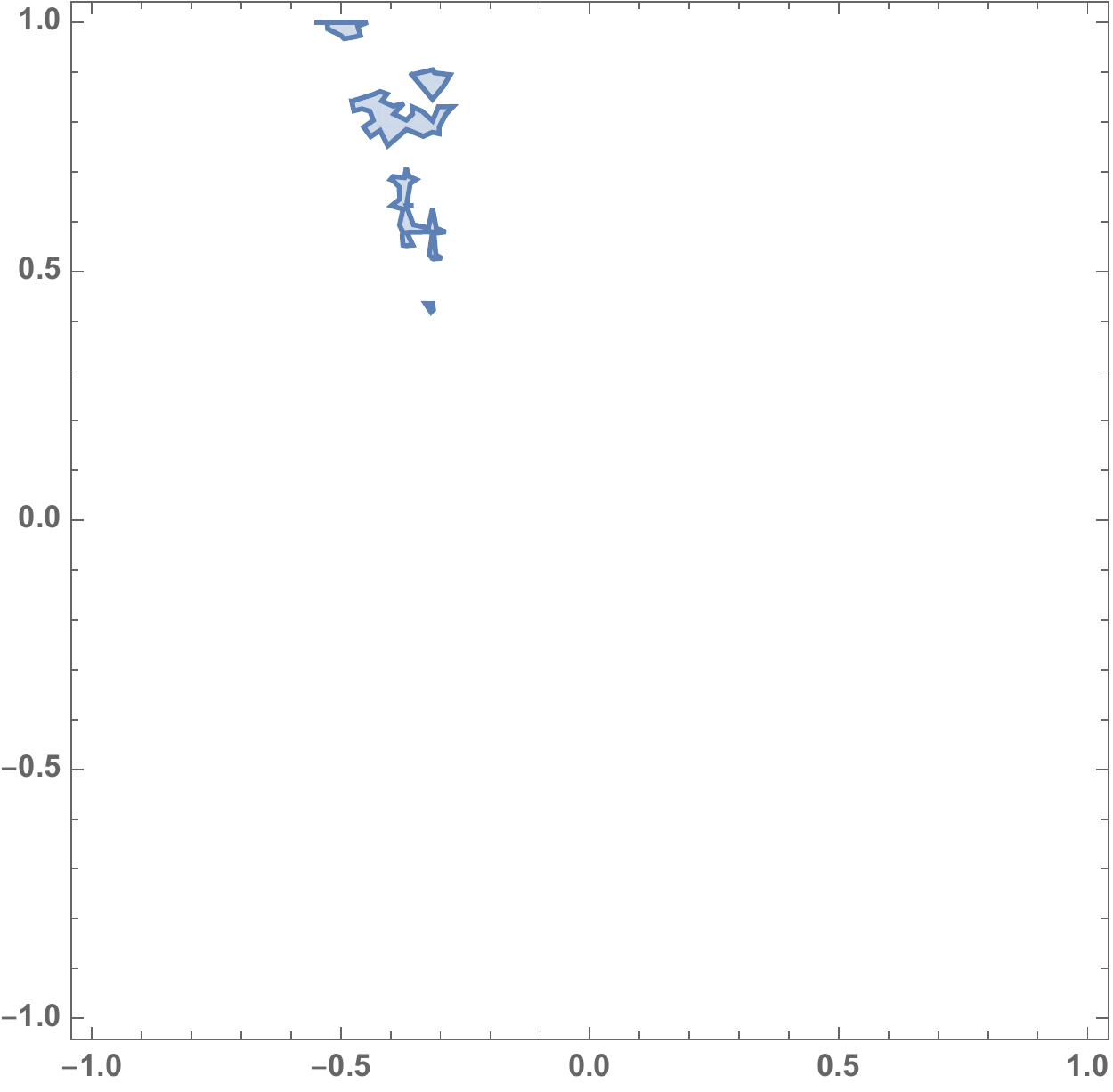} \\ \includegraphics[width=2in]{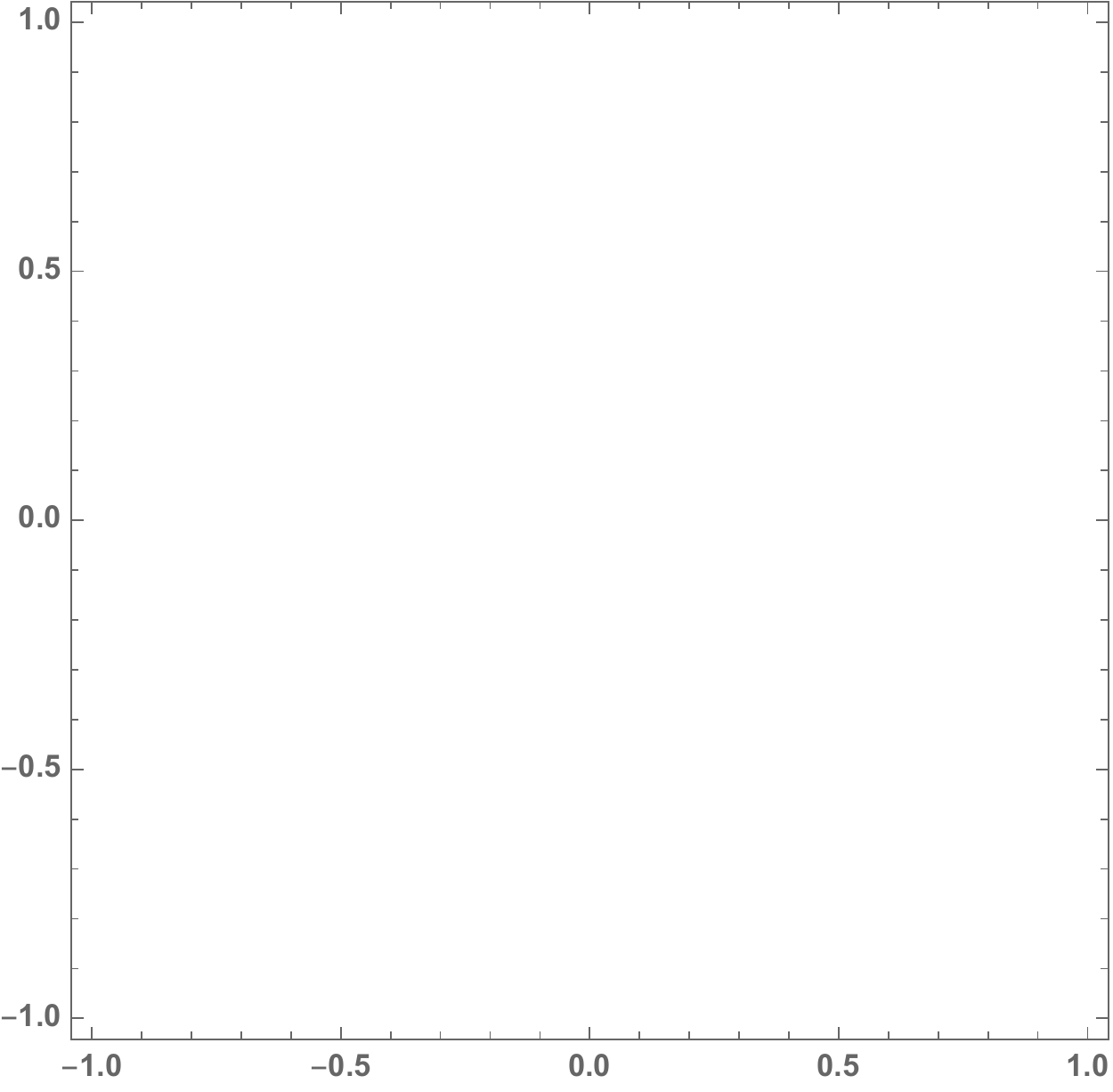} \qquad \includegraphics[width=2in]{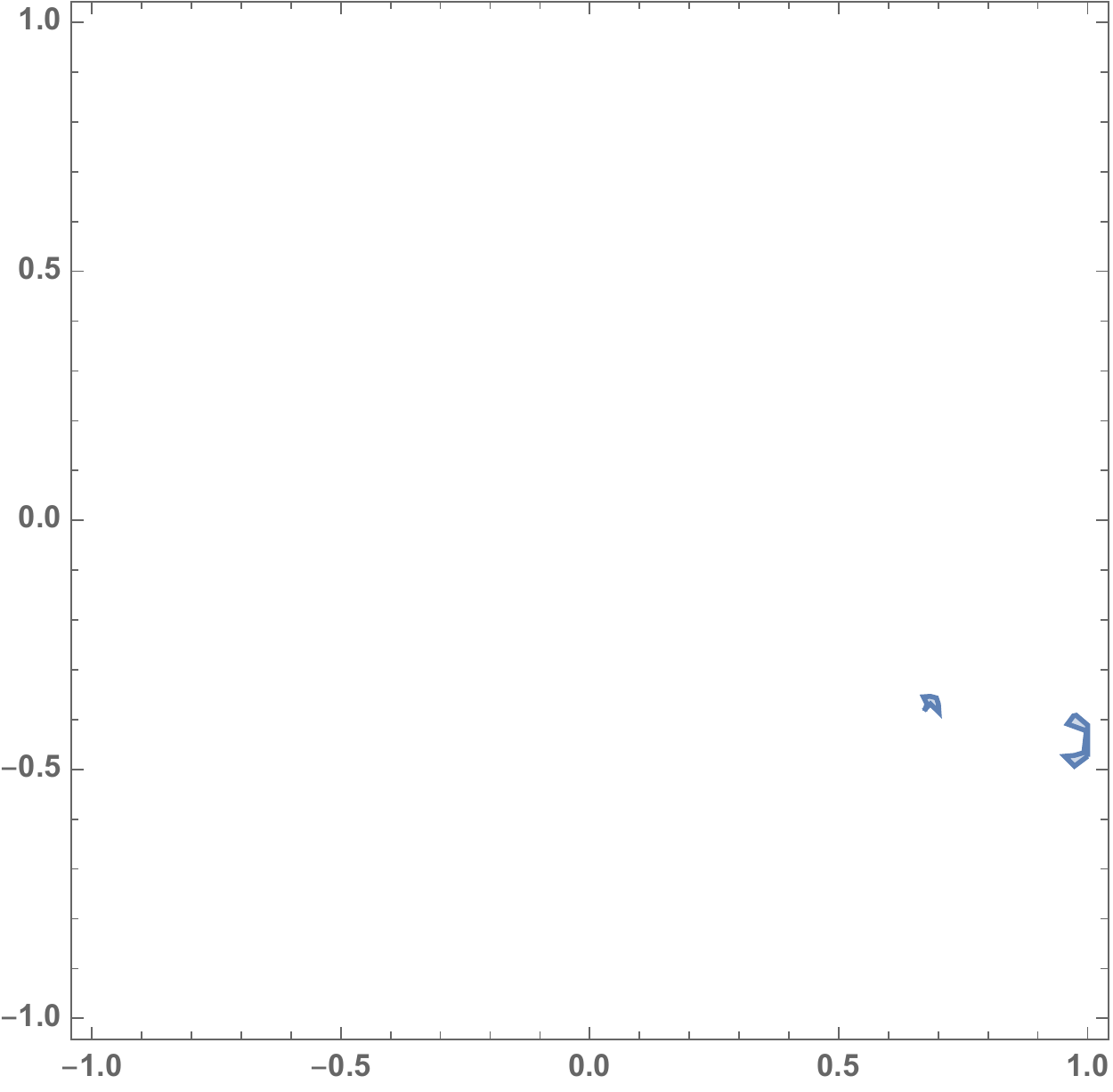}
\caption{{\it Positivity range of $(b_1,b_2)$ for the $H^2\,\times\,H^2$, $H^2\,\times\,S^2$, $S^2\,\times\,H^2$ and $S^2\,\times\,S^2$ backgrounds, from the top-left and clockwise. In the $H^2\,\times\,H^2$ plot, the black dot at $(b_1,b_2)\,=\,(0,0)$ corresponds to the $AdS_2$ solution in \eqref{ads2ads2}.}}
\label{1}
\end{center}
\end{figure}

\subsection{Numerical black hole solutions}

Now we present the full black hole solution numerically. The full black hole solution is an interpolating geometry between the asymptotically $AdS_6$ boundary and the $AdS_2\,\times\,H^2\,\times\,H^2$ horizon. We introduce a new radial coordinate,
\begin{equation}
\rho\,=\,F+\sigma\,.
\end{equation}
This kind of coordinate was introduced in \cite{Benini:2013cda}. Employing the supersymmetry equations, we obtain
\begin{equation}
\frac{\partial\rho}{\partial{r}}\,=\,F'+\sigma'\,=\,-e^FD\,,
\end{equation}
where we define
\begin{equation}
D\,=\,2me^{-3\sigma}+\frac{1}{2m}\left(a_1a_2-b_1b_2\right)e^{\sigma-2G_1-2G_2}\,.
\end{equation}
Then, the supersymmetry equations are
\begin{align}
-D\frac{\partial{F}}{\partial\rho}\,=\,&-\frac{1}{2}\left(g_1e^\sigma\cosh\varphi_2+me^{-3\sigma}\right)-\frac{3}{8m}\left(a_1a_2-b_1b_2\right)e^{\sigma-2G_1-2G_2} \notag \\ &-\frac{\lambda}{4}e^{-\sigma}\cosh\varphi_2\left(a_1e^{-2G_1}+a_2e^{-2G_2}\right)+\frac{\lambda}{4}e^{-\sigma}\sinh\varphi_2\left(b_1e^{-2G_1}+b_2e^{-2G_2}\right)\,, \notag \\
-D\frac{\partial{G_1}}{\partial\rho}\,=\,&-\frac{1}{2}\left(g_1e^\sigma\cosh\varphi_2+me^{-3\sigma}\right)+\frac{1}{8m}\left(a_1a_2-b_1b_2\right)e^{\sigma-2G_1-2G_2} \notag \\ &+\frac{\lambda}{4}e^{-\sigma}\cosh\varphi_2\left(3a_1e^{-2G_1}-a_2e^{-2G_2}\right)-\frac{\lambda}{4}e^{-\sigma}\sinh\varphi_2\left(3b_1e^{-2G_1}-b_2e^{-2G_2}\right)\,, \notag \\
-D\frac{\partial{G_2}}{\partial\rho}\,=\,&-\frac{1}{2}\left(g_1e^\sigma\cosh\varphi_2+me^{-3\sigma}\right)+\frac{1}{8m}\left(a_1a_2-b_1b_2\right)e^{\sigma-2G_1-2G_2} \notag \\ &+\frac{\lambda}{4}e^{-\sigma}\cosh\varphi_2\left(3a_2e^{-2G_2}-a_1e^{-2G_1}\right)-\frac{\lambda}{4}e^{-\sigma}\sinh\varphi_2\left(3b_2e^{-2G_2}-b_1e^{-2G_1}\right)\,, \notag \\
-D\frac{\partial{\sigma}}{\partial\rho}\,=\,&+\frac{1}{2}\left(g_1e^\sigma\cosh\varphi_2-3me^{-3\sigma}\right)-\frac{1}{8m}\left(a_1a_2-b_1b_2\right)e^{\sigma-2G_1-2G_2} \notag \\ &+\frac{\lambda}{4}e^{-\sigma}\cosh\varphi_2\left(a_1e^{-2G_1}+a_2e^{-2G_2}\right)-\frac{\lambda}{4}e^{-\sigma}\sinh\varphi_2\left(b_1e^{-2G_1}+b_2e^{-2G_2}\right)\,, \notag \\
-D\frac{\partial{\varphi_2}}{\partial\rho}\,=\,&+2g_1e^\sigma\sinh\varphi_2-{\lambda}e^{-\sigma}\sinh\varphi_2\left(a_1e^{-2G_1}+a_2e^{-2G_2}\right)+{\lambda}e^{-\sigma}\cosh\varphi_2\left(b_1e^{-2G_1}+b_2e^{-2G_2}\right)\,.
\end{align}
In the $r$-coordinate, the UV or asymptotically $AdS_6$ boundary is at $r\,=\,0$, and the IR or $AdS_2\,\times\,H^2\,\times\,H^2$ horizon is at $r\,=\,\infty$. In this $\rho$-coordinate, the UV is at $\rho\,=\,+\infty$, and the IR is at $\rho\,=\,-\infty$. We present some representative plots of the full black hole solutions in figure 2.
\begin{figure}[h!]
\begin{center}
\includegraphics[width=2.0in]{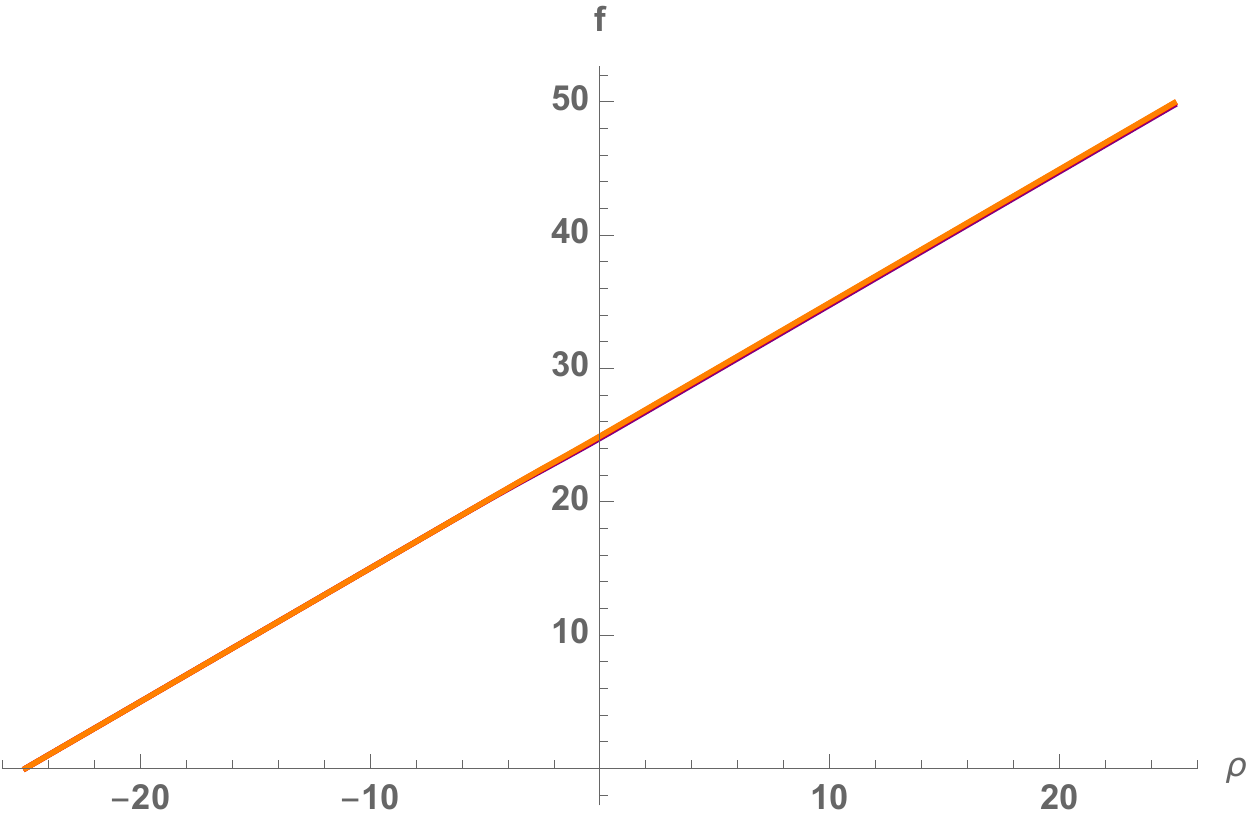} \qquad \includegraphics[width=2.0in]{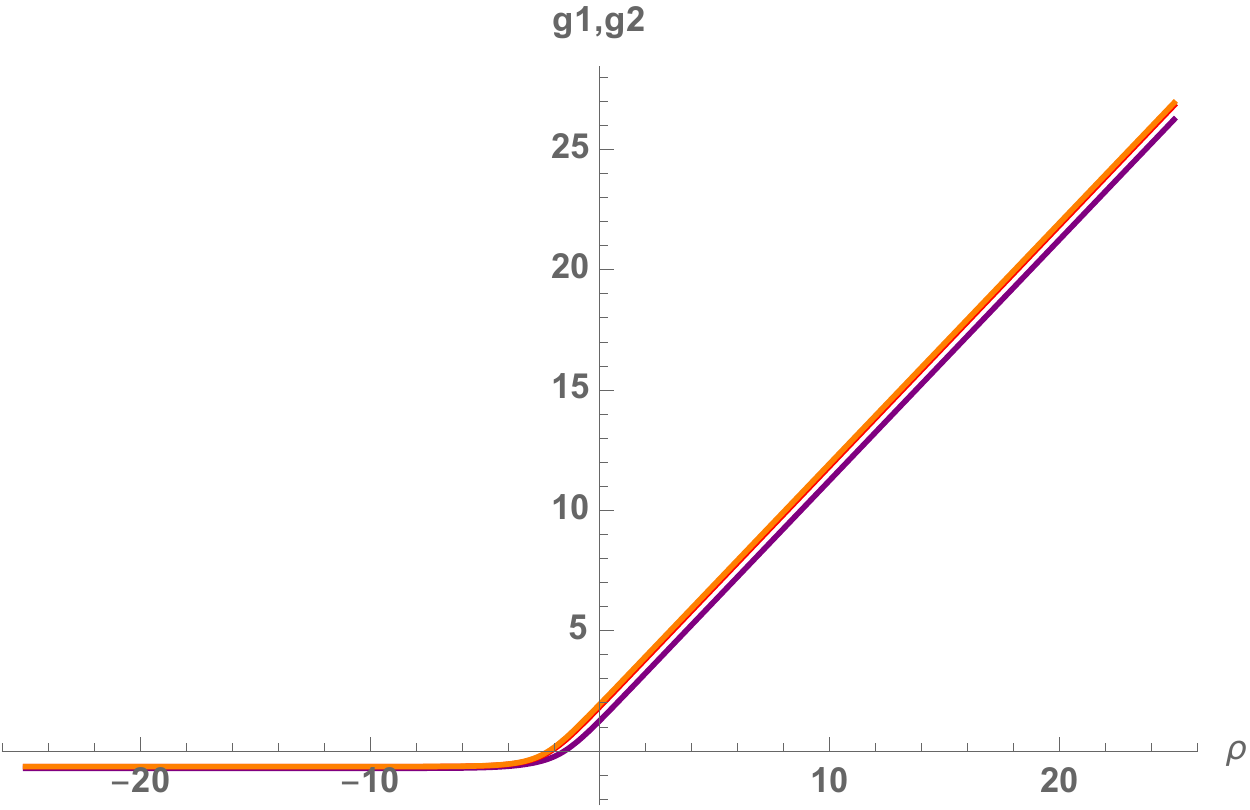} \\ \includegraphics[width=2.0in]{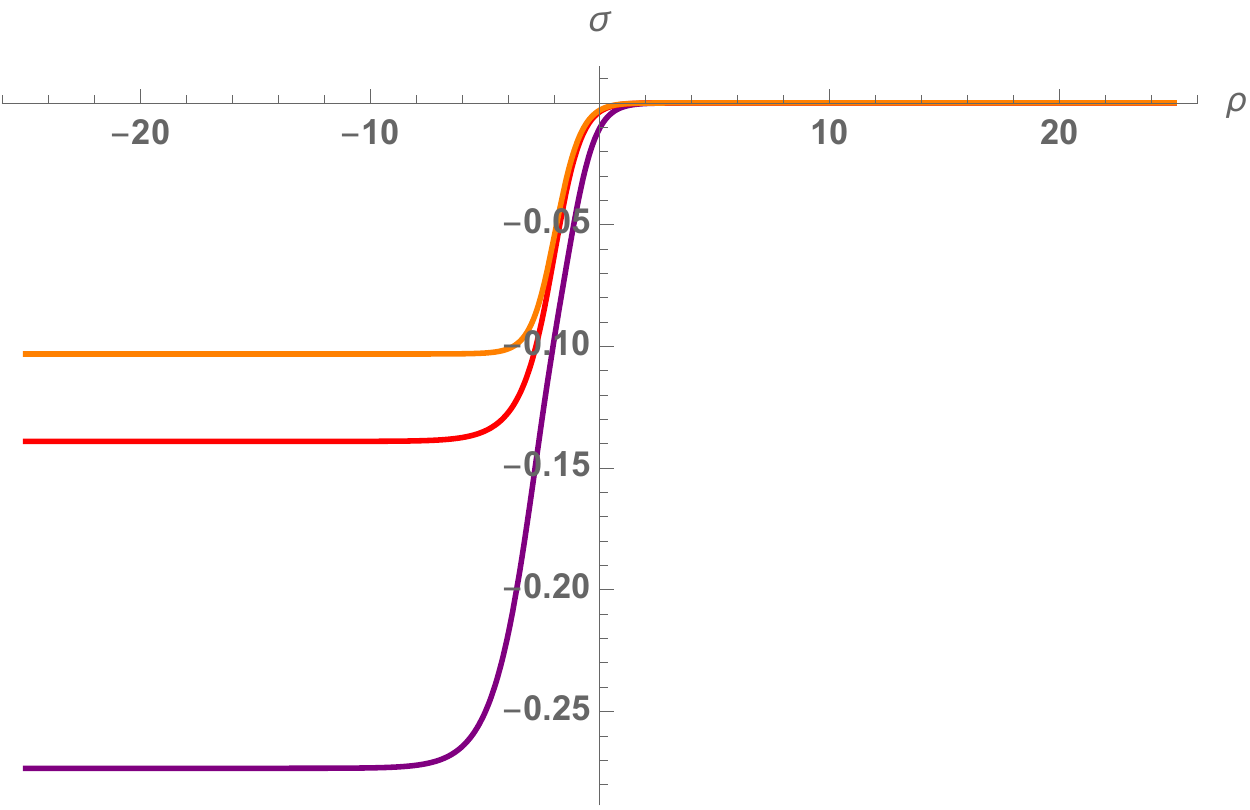} \qquad \includegraphics[width=2.0in]{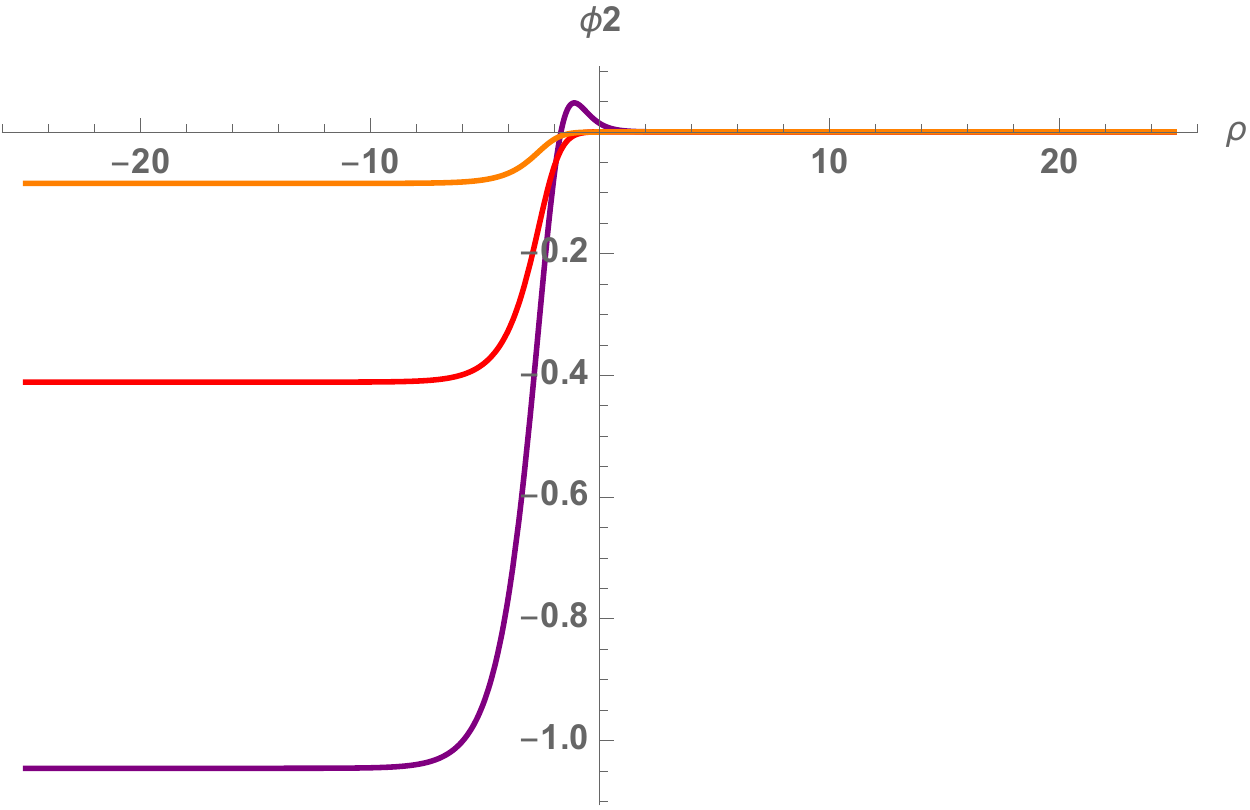}
\caption{{\it Numerical black hole solutions with $m\,=\,1/2$ and $g_1\,=\,3m$. For the magnetic charges, $(b_1,b_2)$, we have $(-0.1,-0.1)$, purple, $(-0.01,-0.01)$, red, and $(-0.001,-0.001)$, orange.}}
\label{1}
\end{center}
\end{figure}

\section{Black holes with other horizons}

In this section, we obtain more black hole solutions with other horizon geometries by considering D4-branes wrapped on K\"ahler four-cycles in Calabi-Yau fourfolds and on Cayley four-cycles in $Spin(7)$ manifolds. We believe these are all possible four-cycles on which D4-branes can wrap in $F(4)$ gauged supergravity. D4-branes on two Riemann surfaces in the previous section fall into a special case of D4-branes on K\"ahler four-cycles in Calabi-Yau fourfolds. The analogous solutions of M5-branes wrapped on supersymmetric four-cycles were studied in \cite{Gauntlett:2000ng, Gauntlett:2001jj}.

\subsection{K\"ahler four-cycles in Calabi-Yau fourfolds}

We consider the $U(1){\times}U(1)$-invariant truncation presented in section 3.1. We consider the metric,
\begin{equation}
ds^2\,=\,e^{2F(r)}\left(dt^2-dr^2\right)-e^{2G(r)}ds^2_{M_4}\,,
\end{equation}
where $M_4$ is a K\"ahler four-cycle in Calabi-Yau fourfolds. The curved coordinates on the K\"ahler four-cycles will be denoted by $\{x_1,\,x_2,\,x_3,\,x_4\}$, and the hatted ones are the flat coordinates. For K\"ahler four-cycles in Calabi-Yau fourfolds, there are four directions transverse to D4-branes in the fourfolds. The normal bundle of the four-cycle has $U(2)\,\subset\,SO(4)$ structure group. We identify $U(1)$ part of the structure group with $U(1)$ gauge field from the non-Abelian $SU(2)$ gauge group, \cite{Gauntlett:2000ng, Gauntlett:1998vk}. The only non-vanishing components of the field strength of $SU(2)$ gauge field, $A^\Lambda_\mu$, $\Lambda\,=\,0,\,1,\,\ldots,\,6$, are given by
\begin{align}
F^3_{\hat{x}_1\hat{x}_2}\,=\,a_1e^{-2G}\,, \qquad F^3_{\hat{x}_3\hat{x}_4}\,=\,a_2e^{-2G}\,, \notag \\
F^6_{\hat{x}_1\hat{x}_2}\,=\,b_1e^{-2G}\,, \qquad F^6_{\hat{x}_3\hat{x}_4}\,=\,b_2e^{-2G}\,,
\end{align}
where the magnetic charges, $a_1$, $a_2$, $b_1$, $b_2$, are constant. The only non-vanishing component of the two-form gauge potential is
\begin{equation}
B_{tr}\,=\,-\frac{1}{2m^2}\left(a_1a_2-b_1b_2\right)e^{2\sigma+2F-4G}\,.
\end{equation}
We employ the projection conditions,
\begin{equation}
\gamma^{\hat{r}}\epsilon_A\,=\,i\epsilon_A\,, \qquad \gamma^{\hat{x}_1\hat{x}_2}\sigma^3_{AB}\epsilon^B\,=\,-i\lambda\epsilon_A\,, \qquad \gamma^{\hat{x}_3\hat{x}_4}\sigma^3_{AB}\epsilon^B\,=\,-i\lambda\epsilon_A\,,
\end{equation}
where $\lambda\,=\,\pm{1}$. Solutions with the projection conditions preserve $1/8$ of the supersymmetries. By employing the projection conditions, we obtain the complete supersymmetry equations,
\begin{align} \label{susyk}
F'e^{-F}\,=\,&-\frac{1}{2}\left(g_1e^\sigma\cosh\varphi_2+me^{-3\sigma}\right)-\frac{\lambda}{2}e^{-\sigma-2G}\left(a\cosh\varphi_2-b\sinh\varphi_2\right)-\frac{3}{8m}\left(a^2-b^2\right)e^{\sigma-4G} \notag \\
G'e^{-F}\,=\,&-\frac{1}{2}\left(g_1e^\sigma\cosh\varphi_2+me^{-3\sigma}\right)+\frac{\lambda}{2}e^{-\sigma-2G}\left(a\cosh\varphi_2-b\sinh\varphi_2\right)+\frac{1}{8m}\left(a^2-b^2\right)e^{\sigma-4G} \notag \\
\sigma'e^{-F}\,=\,&+\frac{1}{2}\left(g_1e^\sigma\cosh\varphi_2-3me^{-3\sigma}\right)+\frac{\lambda}{2}e^{-\sigma-2G}\left(a\cosh\varphi_2-b\sinh\varphi_2\right)-\frac{1}{8m}\left(a^2-b^2\right)e^{\sigma-4G} \notag \\
\varphi_2'e^{-F}\,=\,&+2g_1e^\sigma\sinh\varphi_2-2{\lambda}e^{-\sigma-2G}\left(a\sinh\varphi_2-b\cosh\varphi_2\right)\,.
\end{align}
with the twist conditions,
\begin{equation}
a\,\equiv\,a_1\,=\,a_2\,=\,-\frac{k}{2{\lambda}g_1}\,, \qquad b\,\equiv\,b_1\,=\,b_2\,.
\end{equation}
where $k$ determines the curvature of the K\"ahler four-cycles in Calabi-Yau fourfolds. There is no condition on $b_1$ and $b_2$.

The product of two Riemann surfaces considered in the previous section is a special case of K\"ahler four-cycles in Calabi-Yau fourfolds. When we identify $G\,\equiv\,G_1\,=\,G_2$ in the supersymmetry equations for D4-branes wrapped on two Riemann surfaces, \eqref{susy11}, we obtain the supersymmetry equations here, \eqref{susyk}. By solving the supersymmetry equations, we find the $AdS_2$ fixed point solutions which are identical to the ones obtained in the previous section.

\subsection{Cayley four-cycles in $Spin(7)$ manifolds}

\subsubsection{The $SU(2)$-invariant truncation}

We considered matter coupled $F(4)$ gauged supergravity coupled to three vector multiplets which has $SU(2)_R{\times}SU(2)$ gauge symmetry. In this section we truncate the theory to the $SU(2)_{diag}\,\subset\,SU(2)_R\,\times\,SU(2)$ invariant sector, which was first considered in \cite{Karndumri:2012vh} and again in section 4.1 of \cite{Karndumri:2015eta}. There is one singlet under $SU(2)_{diag}$ which corresponds to $Y_{11}+Y_{22}+Y_{33}$ by the non-compact generators defined in \eqref{generators}. We exponentiate the non-compact generators and obtain the coset representative,
\begin{equation}
L\,=\,e^{\varphi(Y_{11}+Y_{22}+Y_{33})}\,.
\end{equation}
We also have a non-Abelian $SU(2)$ gauge field and a two-form gauge potential in the $SU(2)_{diag}$-invariant truncation. The Lagrangian of the truncation is given by
\begin{align} \label{lagrangian}
e^{-1}\mathcal{L}\,=\,&-\frac{1}{4}R+\partial_\mu\sigma\partial^\mu\sigma+\frac{3}{4}\partial_\mu\varphi\partial^\mu\varphi-V \notag \\
&+\frac{1}{8}\sinh^2(2\varphi)\left[(g_1A^1_\mu-g_2A^4_\mu)^2+(g_1A^2_\mu-g_2A^5_\mu)^2+(g_1A^3_\mu-g_2A^6_\mu)^2\right] \notag \\
&-\frac{1}{8}e^{-2\sigma}\cosh(2\varphi)\left(F^1_{\mu\nu}F^{1\mu\nu}+F^2_{\mu\nu}F^{2\mu\nu}+F^3_{\mu\nu}F^{3\mu\nu}+F^4_{\mu\nu}F^{4\mu\nu}+F^5_{\mu\nu}F^{5\mu\nu}+F^6_{\mu\nu}F^{6\mu\nu}\right) \notag \\ 
&+\frac{1}{4}e^{-2\sigma}\sinh(2\varphi)\left(F^1_{\mu\nu}F^{4\mu\nu}+F^2_{\mu\nu}F^{5\mu\nu}+F^3_{\mu\nu}F^{6\mu\nu}\right)-\frac{1}{8}m^2e^{-2\sigma}B_{\mu\nu}B^{\mu\nu}+\frac{3}{64}e^{4\sigma}H_{\mu\nu\rho}H^{\mu\nu\rho} \notag \\
&-\frac{1}{64}\epsilon^{\mu\nu\rho\sigma\tau\kappa}B_{\mu\nu}\left(F^1_{\rho\sigma}F^1_{\tau\kappa}+F^2_{\rho\sigma}F^2_{\tau\kappa}+F^3_{\rho\sigma}F^3_{\tau\kappa}-F^4_{\rho\sigma}F^4_{\tau\kappa}-F^5_{\rho\sigma}F^5_{\tau\kappa}-F^6_{\rho\sigma}F^6_{\tau\kappa}+\frac{1}{3}m^2B_{\rho\sigma}B_{\tau\kappa}\right)\,,
\end{align}
where the scalar potential is
\begin{align}
V\,=\,\frac{1}{16}g_1^2e^{2\sigma}\left(\cosh(6\varphi)-9\cosh(2\varphi)-8\right)+\frac{1}{16}g_2^2e^{2\sigma}\left(\cosh(6\varphi)-9\cosh(2\varphi)+8\right) \notag \\
-\frac{1}{2}g_1g_2e^{2\sigma}\sinh^3(2\varphi)-4g_1me^{-2\sigma}\cosh^3\varphi+4g_2me^{-2\sigma}\sinh^3\varphi+m^2e^{-6\sigma}\,.
\end{align}

\subsubsection{The supersymmetry equations}

We consider the metric,
\begin{equation}
ds^2\,=\,e^{2F(r)}\left(dt^2-dr^2\right)-e^{2G(r)}ds^2_{M_4}\,,
\end{equation}
where $M_4$ is a Cayley four-cycle in manifolds with $Spin(7)$ holonomy. The curved coordinates on the Cayley four-cycles will be denoted by $\{x_1,\,x_2,\,x_3,\,x_4\}$, and the hatted ones are the flat coordinates. In order to preserve supersymmetry for D4-branes wrapped on Cayley four-cycles in $Spin(7)$ manifolds, we identify self-dual $SU(2)_+$ subgroup of the $SO(4)$ isometry of the four-cycle,
\begin{equation}
SO(4)\,\rightarrow\,SU(2)_+\,\times\,SU(2)_-\,,
\end{equation}
with the non-Abelian $SU(2)$ gauge group, \cite{Gauntlett:2000ng, Gauntlett:1998vk}. The self-duality is defined by
\begin{equation}
\gamma_{\mu\nu}\,=\,\pm\frac{1}{2}\epsilon_{\mu\nu\rho\sigma}\gamma^{\rho\sigma}\,,
\end{equation}
and we denoted the self-duality and anti-self-duality by $+$ and $-$, respectively. For the self-dual part, components are identified by
\begin{equation}
\gamma^{\hat{x}_1\hat{x}_2}\,=\,\gamma^{\hat{x}_3\hat{x}_4}\,, \qquad \gamma^{\hat{x}_1\hat{x}_3}\,=\,\gamma^{\hat{x}_4\hat{x}_2}\,, \qquad \gamma^{\hat{x}_1\hat{x}_4}\,=\,\gamma^{\hat{x}_2\hat{x}_3}\,.
\end{equation}
The only non-vanishing components of the field strength of the $SU(2)$ gauge field, $A^\Lambda_\mu$, $\Lambda\,=\,0,\,1,\,\ldots,\,6$, are given by
\begin{align}
F^1_{\hat{x}_1\hat{x}_2}\,=\,\,F^1_{\hat{x}_3\hat{x}_4}\,=\,a_1e^{-2G}\,, \qquad F^4_{\hat{x}_1\hat{x}_2}\,=\,\,F^4_{\hat{x}_3\hat{x}_4}\,=\,b_1e^{-2G}\,,\notag \\
F^2_{\hat{x}_1\hat{x}_3}\,=\,\,F^2_{\hat{x}_4\hat{x}_2}\,=\,a_2e^{-2G}\,, \qquad F^5_{\hat{x}_1\hat{x}_3}\,=\,\,F^5_{\hat{x}_4\hat{x}_2}\,=\,b_2e^{-2G}\,, \notag \\
F^3_{\hat{x}_1\hat{x}_4}\,=\,\,F^3_{\hat{x}_2\hat{x}_3}\,=\,a_3e^{-2G}\,, \qquad F^6_{\hat{x}_1\hat{x}_4}\,=\,\,F^6_{\hat{x}_2\hat{x}_3}\,=\,b_3e^{-2G}\,,
\end{align}
where the magnetic charges, $a_1$, $a_2$, $a_3$, $b_1$, $b_2$, $b_3$, are constant. As we have one $SU(2)_{diag}\,\subset\,SU(2)_R\,\times\,SU(2)$ non-Abelian gauge field, they are related by
\begin{equation} \label{gagb}
g_1a_1\,=\,g_2b_1\,, \qquad g_1a_2\,=\,g_2b_2\,, \qquad g_1a_3\,=\,g_2b_3\,,
\end{equation}
where $g_1$ and $g_2$ are the gauge coupling constants of $SU(2)_R$ and $SU(2)$, respectively. The only non-vanishing component of the two-form gauge potential is
\begin{equation}
B_{tr}\,=\,-\frac{1}{2m^2}\left(a_1^2+a_2^2+a_3^2-b_1^2-b_2^2-b_3^2\right)e^{2\sigma+2F-4G}\,.
\end{equation}
We employ the projection conditions,
\begin{align}
\gamma^{\hat{r}}&\epsilon_A\,=\,i\epsilon_A\,, \notag \\
\gamma^{\hat{x}_1\hat{x}_2}\sigma^1_{AB}\epsilon^B\,=&\,\gamma^{\hat{x}_3\hat{x}_4}\sigma^1_{AB}\epsilon^B\,=\,-i\lambda\epsilon_A\,, \notag \\
\gamma^{\hat{x}_1\hat{x}_3}\sigma^2_{AB}\epsilon^B\,=&\,\gamma^{\hat{x}_4\hat{x}_2}\sigma^2_{AB}\epsilon^B\,=\,-i\lambda\epsilon_A\,, \notag \\
\gamma^{\hat{x}_1\hat{x}_4}\sigma^3_{AB}\epsilon^B\,=&\,\gamma^{\hat{x}_2\hat{x}_3}\sigma^3_{AB}\epsilon^B\,=\,-i\lambda\epsilon_A\,,
\end{align}
where $\lambda\,=\,\pm{1}$. Solutions with the projection conditions preserve $1/16$ of the supersymmetries. 

Employing the projection conditions, from $\delta\lambda_A^I\,=\,0$, we obtain  for $I\,=\,1,\,2,\,3$, respectively,
\begin{align}
\varphi'e^{-F}\,=\,&+(g_1e^\sigma\cosh\varphi-g_2e^\sigma\sinh\varphi)\sinh(2\varphi)-2{\lambda}e^{-\sigma-2G}\left(a_1\sinh\varphi-b_1\cosh\varphi\right)\,, \notag \\
\varphi'e^{-F}\,=\,&+(g_1e^\sigma\cosh\varphi-g_2e^\sigma\sinh\varphi)\sinh(2\varphi)-2{\lambda}e^{-\sigma-2G}\left(a_2\sinh\varphi-b_2\cosh\varphi\right)\,, \notag \\
\varphi'e^{-F}\,=\,&+(g_1e^\sigma\cosh\varphi-g_2e^\sigma\sinh\varphi)\sinh(2\varphi)-2{\lambda}e^{-\sigma-2G}\left(a_3\sinh\varphi-b_3\cosh\varphi\right)\,.
\end{align}
Therefore, we conclude that the magnetic charges are
\begin{equation}
a\,\equiv\,a_1\,=\,a_2\,=a_3\,, \qquad b\,\equiv\,b_1\,=\,b_2\,=b_3\,.
\end{equation}

We present the complete supersymmetry equations,
\begin{align}
F'e^{-F}\,=\,&-\frac{1}{2}\left(g_1e^\sigma\cosh^3\varphi-g_2e^\sigma\sinh^3\varphi+me^{-3\sigma}\right)-\frac{3\lambda}{2}e^{-\sigma-2G}\left(a\cosh\varphi-b\sinh\varphi\right) \notag \\ 
&-\frac{9}{8m}\left(a^2-b^2\right)e^{\sigma-4G}\,, \notag \\
G'e^{-F}\,=\,&-\frac{1}{2}\left(g_1e^\sigma\cosh^3\varphi-g_2e^\sigma\sinh^3\varphi+me^{-3\sigma}\right)+\frac{3\lambda}{2}e^{-\sigma-2G}\left(a\cosh\varphi-b\sinh\varphi\right) \notag \\ 
&+\frac{3}{8m}\left(a^2-b^2\right)e^{\sigma-4G}\,, \notag \\
\sigma'e^{-F}\,=\,&+\frac{1}{2}\left(g_1e^\sigma\cosh^3\varphi-g_2e^\sigma\sinh^3\varphi-3me^{-3\sigma}\right)+\frac{3\lambda}{2}e^{-\sigma-2G}\left(a\cosh\varphi-b\sinh\varphi\right) \notag \\ 
&-\frac{3}{8m}\left(a^2-b^2\right)e^{\sigma-4G}\,, \notag \\
\varphi'e^{-F}\,=\,&+(g_1e^\sigma\cosh\varphi-g_2e^\sigma\sinh\varphi)\sinh(2\varphi)-2{\lambda}e^{-\sigma-2G}\left(a\sinh\varphi-b\cosh\varphi\right)\,.
\end{align}
We also obtain twist conditions on the magnetic charges,
\begin{equation} \label{ctwist}
a\,=\,-\frac{k}{6{\lambda}g_1}\,, \qquad b\,=\,-\frac{k}{6{\lambda}g_2}\,,
\end{equation}
where $k$ determines the curvature of the Cayley four-cycles in $Spin(7)$ manifolds. The twist condition on $b$ comes from the $SU(2)_{diag}$ condition, \eqref{gagb}.

\subsubsection{The $AdS_2$ solutions}

Now  we will consider the $\mathcal{N}\,=\,4^+$ theory, $g_1>0$, $m>0$. When $b\,=\,0$, we find an $AdS_2$ fixed point solution for $k\,=\,-1$,
\begin{equation}
e^{F}\,=\,\frac{3^{1/4}}{2^{3/2}g_1^{3/4}m^{1/4}}\frac{1}{r}\,, \qquad e^{G}\,=\,\frac{1}{2^{1/2}3^{1/4}g_1^{3/4}m^{1/4}}\,, \qquad e^{\sigma}\,=\,\frac{2^{1/2}m^{1/4}}{3^{1/4}g_1^{1/4}}\,, \qquad e^{\varphi}\,=\,1\,.
\end{equation}
After the reparametrization of the parameters given in \eqref{reparam}, this is the $AdS_2$ solution first found in \cite{Suh:2018tul}.

When we consider for non-zero $b$, we find new $AdS_2$ solutions in terms of the scalar field, $\varphi$,
\begin{align} \label{cAdS21}
e^F\,=&\,\frac{e^{-\sigma}}{2(g_1\cosh^3\varphi-g_2\sinh^3\varphi)}\frac{1}{r}\,, \notag \\
e^{2G}\,=&\,\frac{3\lambda}{2m}e^{2\sigma}(a\cosh\varphi-b\sinh\varphi)\,, \notag \\
e^{4\sigma}\,=&\,\frac{m}{6}\frac{5(a^2-b^2)+3(a^2+b^2)\cosh(2\varphi)-6ab\sinh(2\varphi)}{(g_1\cosh^3\varphi-g_2\sinh^3\varphi)(a\cosh\varphi-b\sinh\varphi)^2}\,.
\end{align}
Then, the scalar field, $\varphi$, should be expressed in terms of the magnetic charges, $a$ and $b$, but the expression is very unwieldy. Alternatively, we present the magnetic charge, $b$, in terms of the scalar field, $\varphi$, 
\begin{align} \label{cAdS22}
b\,=&\,-a\Big[g_1\left(1+e^{2\varphi}\right)^3\left(1-6e^{2\varphi}+e^{4\varphi}\right)+g_2\left(1-e^{2\varphi}\right)^3\left(1+6e^{2\varphi}+e^{4\varphi}\right) \notag \\ 
+&4e^{2\varphi}\sqrt{g_1^2\left(1+e^{2\varphi}\right)^4\left(5-6e^{2\varphi}+5e^{4\varphi}\right)+g_2^2\left(1-e^{2\varphi}\right)^4\left(5+6e^{2\varphi}+5e^{4\varphi}\right)+10g_1g_2\left(1-e^{4\varphi}\right)^3}\Big] \notag \\ 
/&\left[\left(1-e^{4\varphi}\right)\left(g_1\left(1-13e^{2\varphi}-13e^{4\varphi}+e^{6\varphi}\right)+g_2\left(1-7e^{2\varphi}+7e^{4\varphi}-e^{6\varphi}\right)\right)\right]\,.
\end{align}
The solutions are parametrized by two magnetic charges, $a$ and $b$. It will be interesting to have a  field theory interpretation of this $AdS_2$ fixed point solution.

Unlike the black holes with a horizon of two Riemann surfaces, for this case, we could not device a way to determine the positivity range for $AdS_2$ solutions. However, as we see in the next subsection, we obtained a number of $AdS_2$ solutions with negative curvature horizon, $k\,=\,-1$, numerically. On the other hand, we could not find any solutions with positive curvature horizon, $k\,=\,+1$. Thus, we will concentrate on solutions with negative curvature horizon, $k\,=\,-1$.

\subsubsection{Numerical black hole solutions}

Now we present the full black hole solution numerically. The full black hole solution is an interpolating geometry between the asymptotically $AdS_6$ boundary and the $AdS_2\,\times\,Cayley_4$ horizon. As we explained at the end of the last subsection, we will concentrate on solutions with negative curvature horizon, $k\,=\,-1$. We introduce a new radial coordinate,
\begin{equation}
\rho\,=\,F+\sigma\,.
\end{equation}
This kind of coordinate was introduced in \cite{Benini:2013cda}. Employing the supersymmetry equations, we obtain
\begin{equation}
\frac{\partial\rho}{\partial{r}}\,=\,F'+\sigma'\,=\,-e^FD\,,
\end{equation}
where we define
\begin{equation}
D\,=\,2me^{-3\sigma}+\frac{3}{2m}\left(a^2-b^2\right)e^{\sigma-4G}\,.
\end{equation}
Then, the supersymmetry equations are
\begin{align}
-D\frac{\partial{F}}{\partial\rho}\,=\,&-\frac{1}{2}\left(g_1e^\sigma\cosh^3\varphi-g_2e^\sigma\sinh^3\varphi+me^{-3\sigma}\right)-\frac{3\lambda}{2}e^{-\sigma-2G}\left(a\cosh\varphi-b\sinh\varphi\right) \notag \\ 
&-\frac{9}{8m}\left(a^2-b^2\right)e^{\sigma-4G}\,, \notag \\
-D\frac{\partial{G}}{\partial\rho}\,=\,&-\frac{1}{2}\left(g_1e^\sigma\cosh^3\varphi-g_2e^\sigma\sinh^3\varphi+me^{-3\sigma}\right)+\frac{3\lambda}{2}e^{-\sigma-2G}\left(a\cosh\varphi-b\sinh\varphi\right) \notag \\ 
&+\frac{3}{8m}\left(a^2-b^2\right)e^{\sigma-4G}\,, \notag \\
-D\frac{\partial{\sigma}}{\partial\rho}\,=\,&+\frac{1}{2}\left(g_1e^\sigma\cosh^3\varphi-g_2e^\sigma\sinh^3\varphi-3me^{-3\sigma}\right)+\frac{3\lambda}{2}e^{-\sigma-2G}\left(a\cosh\varphi-b\sinh\varphi\right) \notag \\ 
&-\frac{3}{8m}\left(a^2-b^2\right)e^{\sigma-4G}\,, \notag \\
-D\frac{\partial{\varphi_2}}{\partial\rho}\,=\,&+(g_1e^\sigma\cosh\varphi-g_2e^\sigma\sinh\varphi)\sinh(2\varphi)-2{\lambda}e^{-\sigma-2G}\left(a\sinh\varphi-b\cosh\varphi\right)\,.
\end{align}
In the $r$-coordinate, the UV or asymptotically $AdS_6$ boundary is at $r\,=\,0$, and the IR or $AdS_2\,\times\,Cayley_4$ horizon is at $r\,=\,\infty$. In this $\rho$-coordinate, the UV is at $\rho\,=\,+\infty$, and the IR is at $\rho\,=\,-\infty$. We set $m\,=\,1/2$ and $g_1\,=\,3m$ to have a unit radius for the $AdS_6$ boundary. Then, $a$ is determined by \eqref{ctwist}, and there is one free parameter left, $b$. Employing \eqref{cAdS21} and \eqref{cAdS22} to determine boundary conditions, we solve the supersymmetry equations numerically. With some choices of $b$, we present representative plots of the full black hole solutions in figure 3.

\begin{figure}[h!]
\begin{center}
\includegraphics[width=2.0in]{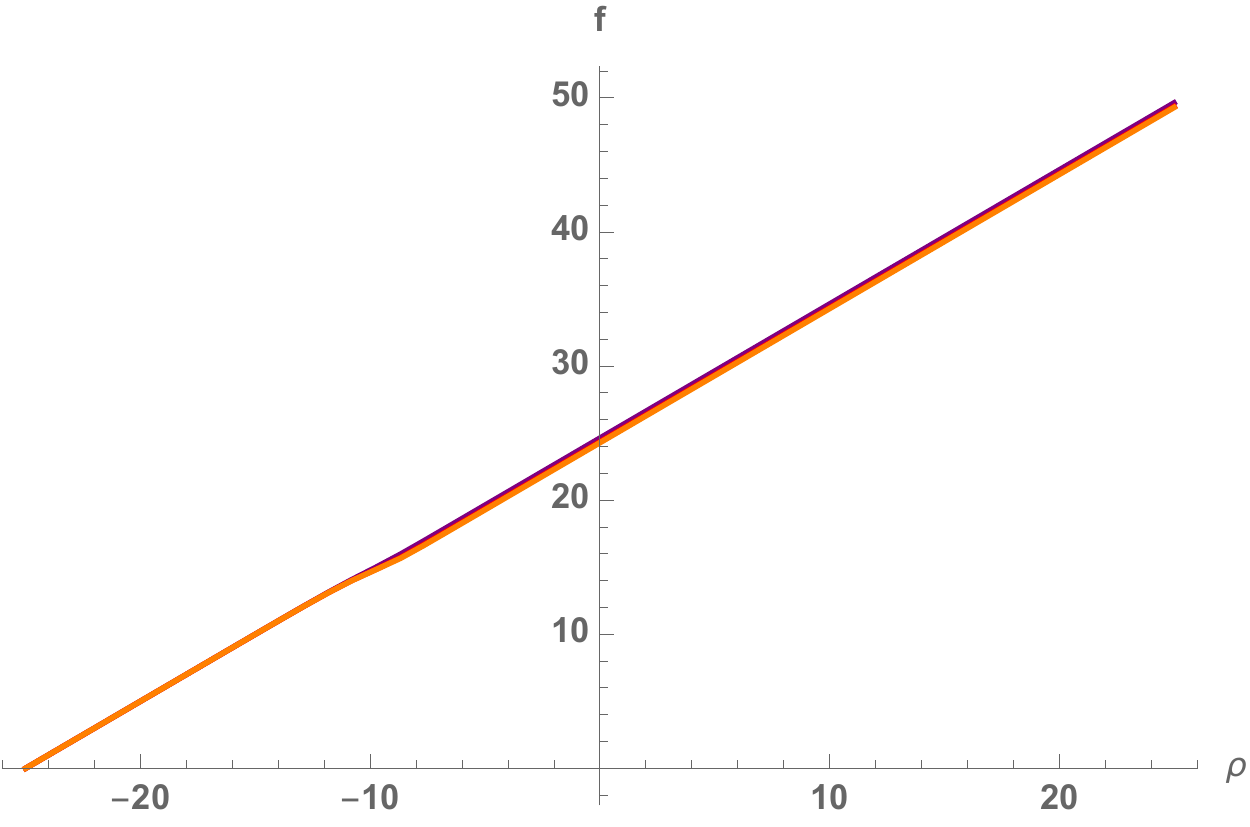} \qquad \includegraphics[width=2.0in]{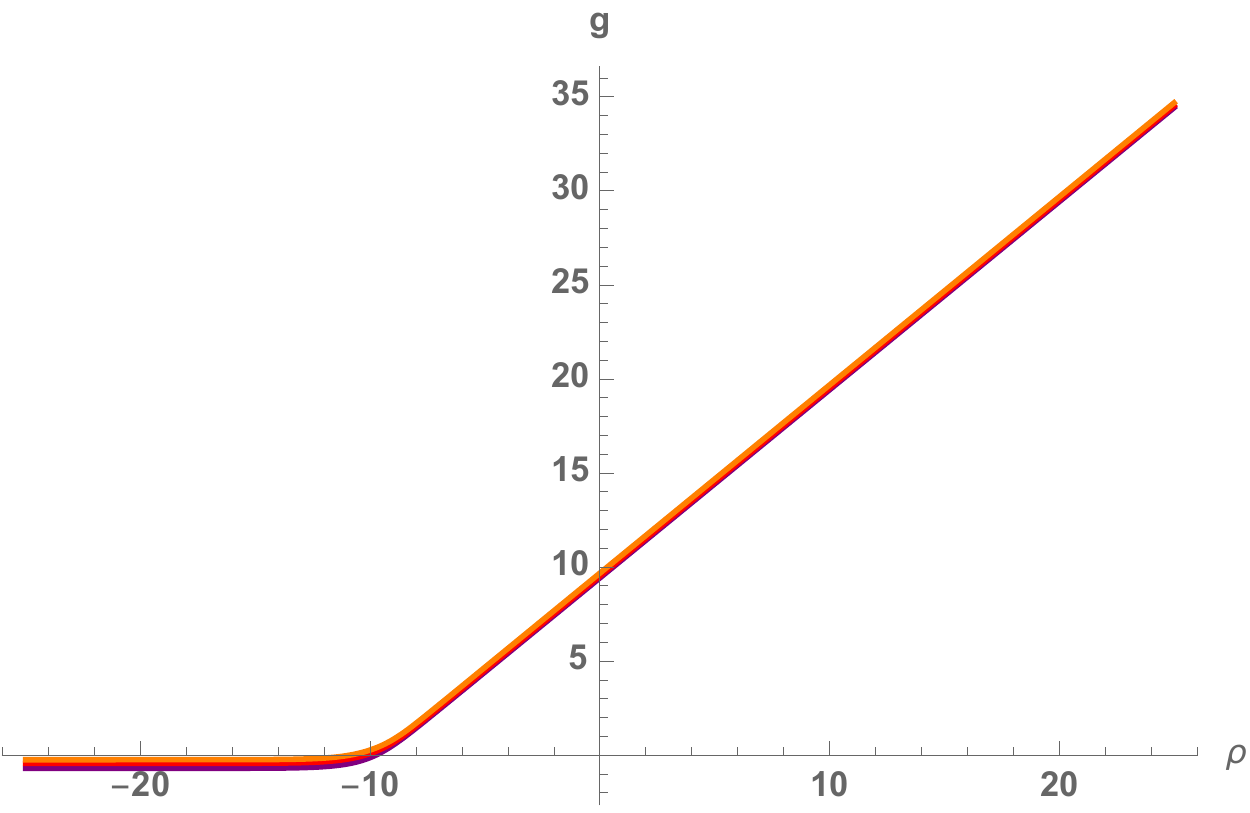} \\ \includegraphics[width=2.0in]{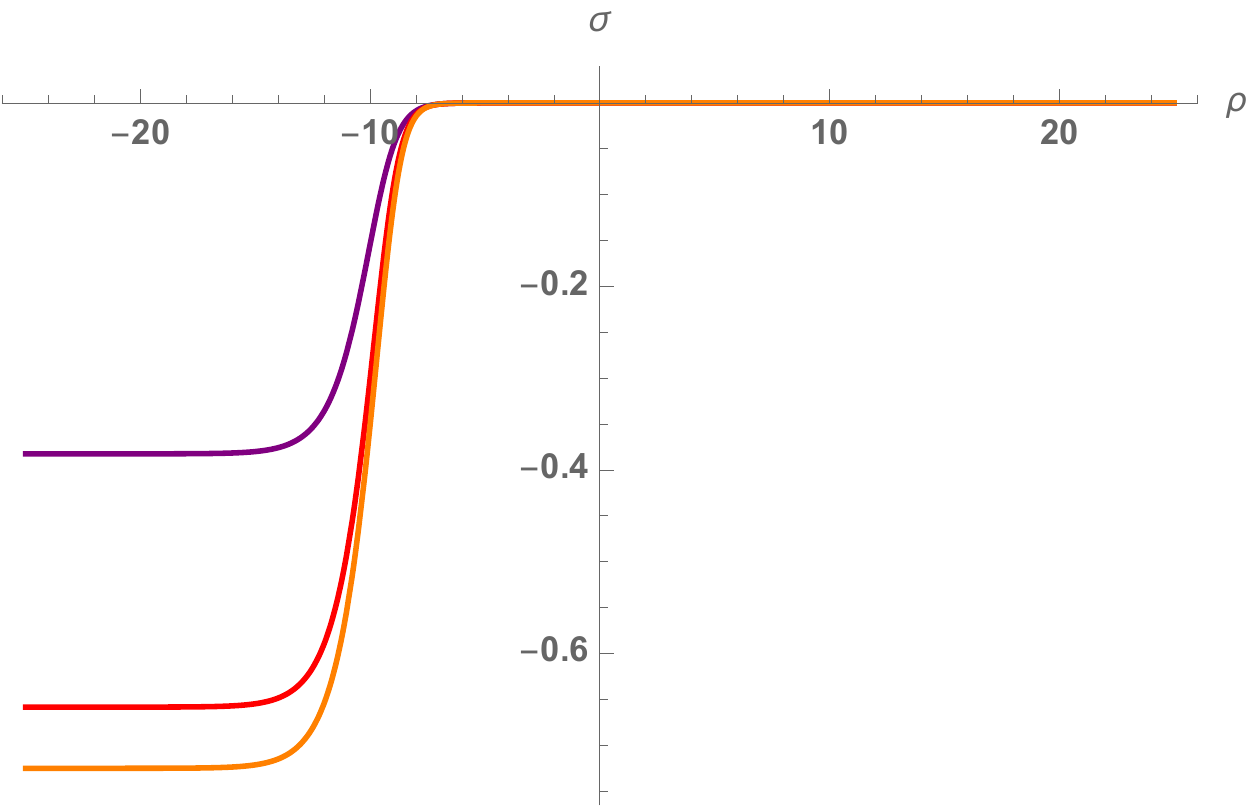} \qquad \includegraphics[width=2.0in]{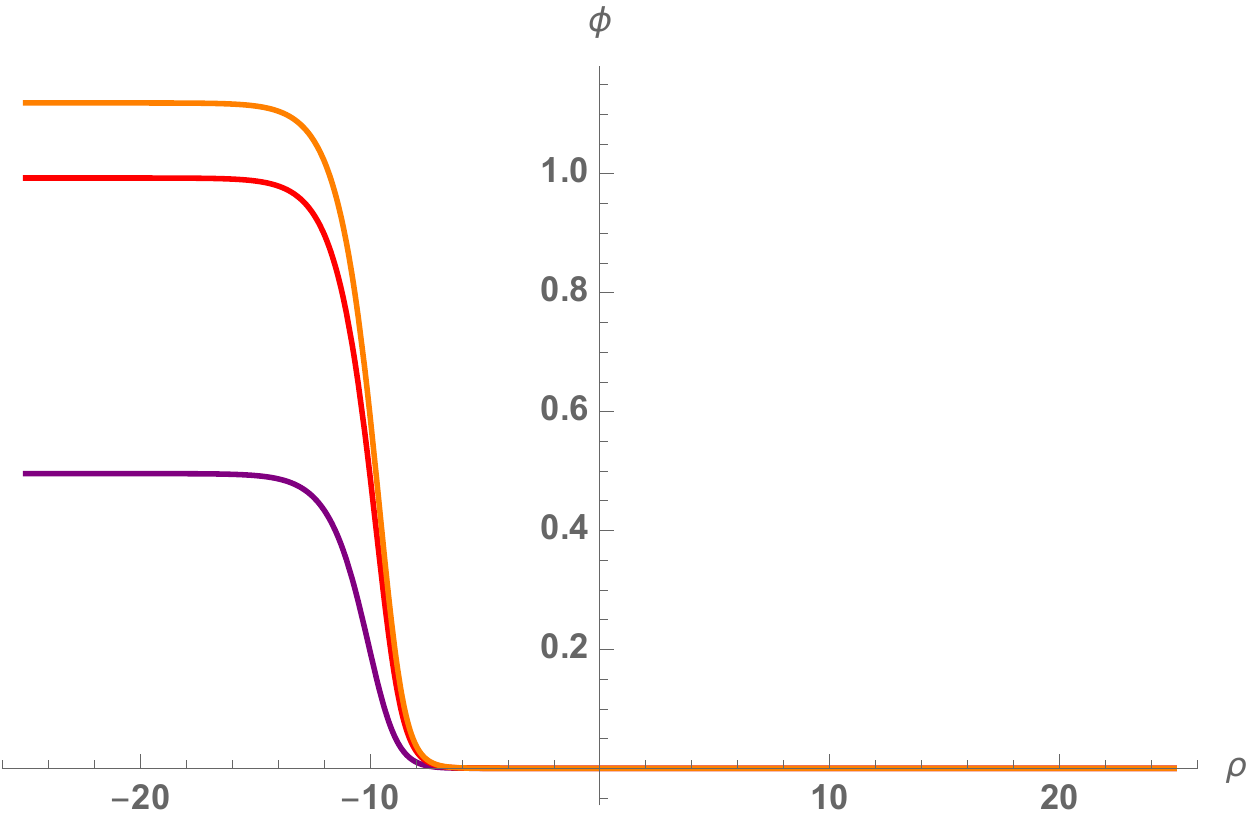}
\caption{{\it Numerical black hole solutions with $m\,=\,1/2$ and $g\,=\,3m$. We have $b=-0.1$ for purple, $b=-0.3$ for red, and $b=-0.5$ for orange.}}
\label{1}
\end{center}
\end{figure}

\bigskip
\bigskip
\bigskip
\leftline{\bf Acknowledgements}
\noindent We would like to thank Parinya Karndumri for helpful discussions. This research was supported by the National Research Foundation of Korea under the grant NRF-2017R1D1A1B03034576.

\appendix
\section{The equations of motion}
\renewcommand{\theequation}{A.\arabic{equation}}
\setcounter{equation}{0} 

In this appendix, we present the equations of motion for the $U(1){\times}U(1)$-invariant truncation in \eqref{lagrangian} with $\varphi_1\,=\,0$ as in \eqref{phi1zero},
\begin{align}
R_{\mu\nu}&-\frac{1}{2}Rg_{\mu\nu}=2\left[Vg_{\mu\nu}+2\left(\partial_\mu\sigma\partial_\nu\sigma-\frac{1}{2}g_{\mu\nu}g^{\rho\sigma}\partial_\rho\sigma\partial_\sigma\sigma\right)+\frac{1}{2}\left(\partial_\mu\varphi_2\partial_\nu\varphi_2-\frac{1}{2}g_{\mu\nu}g^{\rho\sigma}\partial_\rho\varphi_2\partial_\sigma\varphi_2\right)\right. \notag \\ 
&-\frac{1}{2}e^{-2\sigma}\cosh(2\varphi_2)\left(g^{\rho\sigma}F^3_{\mu\rho}F^3_{\nu\sigma}-\frac{1}{4}g_{\mu\nu}F^3_{\rho\sigma}F^{3\rho\sigma}\right)-\frac{1}{2}e^{-2\sigma}\cosh(2\varphi_2)\left(g^{\rho\sigma}F^6_{\mu\rho}F^6_{\nu\sigma}-\frac{1}{4}g_{\mu\nu}F^6_{\rho\sigma}F^{6\rho\sigma}\right) \notag \\
&+e^{-2\sigma}\sinh(2\varphi_2)\left(g^{\rho\sigma}F^3_{\mu\rho}F^6_{\nu\sigma}-\frac{1}{4}g_{\mu\nu}F^3_{\rho\sigma}F^{6\rho\sigma}\right)\left.-\frac{1}{2}m^2e^{-2\sigma}\left(g^{\rho\sigma}B_{\mu\rho}B_{\nu\sigma}-\frac{1}{4}g_{\mu\nu}B_{\rho\sigma}B^{\rho\sigma}\right)\right]\,,
\end{align}
\begin{align}
\frac{1}{\sqrt{-g}}\partial_\mu\left(\sqrt{-g}g^{\mu\nu}\partial_\nu\sigma\right)&+\frac{1}{2}\frac{\partial{V}}{\partial\sigma}-\frac{1}{8}e^{-2\sigma}\cosh(2\varphi_2)F^3_{\mu\nu}F^{3\mu\nu}-\frac{1}{8}e^{-2\sigma}\cosh(2\varphi_2)F^6_{\mu\nu}F^{6\mu\nu} \notag \\ &+\frac{1}{4}e^{-2\sigma}\cosh(2\varphi_2)F^3_{\mu\nu}F^{6\mu\nu}-\frac{1}{8}m^2e^{-2\sigma}B_{\mu\nu}B^{\mu\nu}\,=\,0\,,
\end{align}
\begin{align}
\frac{1}{\sqrt{-g}}\partial_\mu\left(\sqrt{-g}g^{\mu\nu}\partial_\nu\varphi_2\right)&+2\frac{\partial{V}}{\partial\varphi_2}+\frac{1}{2}e^{-2\sigma}\cosh(2\varphi_2)F^3_{\mu\nu}F^{3\mu\nu}+\frac{1}{2}e^{-2\sigma}\cosh(2\varphi_2)F^6_{\mu\nu}F^{6\mu\nu} \notag \\ &-e^{-2\sigma}\cosh(2\varphi_2)F^3_{\mu\nu}F^{6\mu\nu}\,=\,0\,.
\end{align}
\begin{align}
&\mathcal{D}_\nu\left(e^{-2\sigma}F^{\Lambda{\nu\mu}}\right)\,=\,\frac{1}{24}e\epsilon^{\mu\nu\rho\sigma\tau\kappa}F^\Lambda_{\nu\rho}H_{\sigma\tau\kappa}\,, \\
&\mathcal{D}_\rho\left(e^{4\sigma}H^{\rho\mu\nu}\right)\,=\,-\frac{1}{16}e\epsilon^{\mu\nu\rho\sigma\tau\kappa}\eta_{\Lambda\Sigma}F^\Lambda_{\rho\sigma}F^\Sigma_{\tau\kappa}-me^{-2\sigma}\delta^{0\Lambda}F^{\Lambda\mu\nu}\,.
\end{align}

\bigskip
\bigskip
\bigskip
\bigskip
\bigskip



\end{document}